\documentclass[12pt]{article}
\hoffset= - 1.7cm
\usepackage{amssymb,graphicx}
\usepackage{epstopdf}
\usepackage{amsmath,amsfonts}
\usepackage{epsfig}
\usepackage[dvipsnames]{xcolor}
\def\const{\mbox{const}}

\def\eps{\epsilon}

\newcommand{\be}{\begin{equation}}
\newcommand{\ee}{\end{equation}}
\newcommand{\bea}{\begin{eqnarray}}
\newcommand{\eea}{\end{eqnarray}}
\newcommand{\bg}{\begin{gather}}
\newcommand{\eg}{\end{gather}}
\newcommand{\bseq}{\begin{subequations}}
\newcommand{\eseq}{\end{subequations}}

\renewcommand{\ln}{\mathop{\rm ln}\nolimits}

\usepackage{hyperref}
\definecolor{linkcolor}{HTML}{799B03}
\definecolor{urlcolor}{HTML}{799B03}
\hypersetup{pdfstartview=FitH,linkcolor=linkcolor,urlcolor=urlcolor,colorlinks=true}

\makeatletter
\newcommand*{\myfnsymbolsingle}[1]{%
  \ensuremath{%
    \ifcase#1
    \or 
      *%
    \or 
      \dagger
    \or 
      \ddagger
    \or 
      1
    \or 
      2
    \or   
      3
    \or
      4
    \or 
      5
    \or
      6
    \or   
      7
    \or
      8
    \else 
      \@ctrerr  
    \fi
  }%
}   
\makeatother

\usepackage{alphalph}
\newalphalph{\myfnsymbolmult}[mult]{\myfnsymbolsingle}{}
\renewcommand*{\thefootnote}{%
  \myfnsymbolmult{\value{footnote}}%
}

\begin{document}

\begin{center}
{\LARGE \bf  Cosmological scenarios with bounce and Genesis in Horndeski theory and beyond}

\vspace{10pt}

{\bf  An essay in honor of I.M. Khalatnikov on the occasion of his 100th birthday}

\vspace{20pt}
S. Mironov$^{a,c,d}$\footnote{sa.mironov\_1@physics.msu.ru},
V. Rubakov$^{a,b}$\footnote{rubakov@inr.ac.ru},
V. Volkova$^{a}$\footnote{volkova.viktoriya@physics.msu.ru}
\renewcommand*{\thefootnote}{\arabic{footnote}}
\vspace{15pt}

$^a$\textit{Institute for Nuclear Research of the Russian Academy of Sciences,\\
60th October Anniversary Prospect, 7a, 117312 Moscow, Russia}\\
\vspace{5pt}

$^b$\textit{Department of Particle Physics and Cosmology, Physics Faculty,\\
M.V. Lomonosov Moscow State University,\\
Vorobjevy Gory, 119991 Moscow, Russia}

$^c$\textit{Institute for Theoretical and Experimental Physics,\\
Bolshaya Cheriomyshkinskaya, 25, 117218 Moscow, Russia}

$^d$\textit{Moscow Institute of Physics and Technology,\\
Institutski pereulok, 9, 141701, Dolgoprudny, Russia}
\end{center}

\vspace{5pt}

\begin{abstract}
This essay is a brief review of the recent studies of non-singular 
cosmological scenarios with bounce and Genesis and their stability in a 
subclass of scalar-tensor theories with higher derivatives 
-- beyond Horndeski theories. We discuss the general results of stability 
analysis of the non-singular cosmological solutions  in beyond 
Horndeski theories, as well as other closely related topics: 1) the
no-go theorem, which is valid in the general Horndeski theories but 
not in their extensions, 2) singularities in disformal 
transformations relating beyond Horndeski theories with general ones,
3) healthy behaviour of the scalar sector in the unitary gauge despite 
divergencies of  coefficients in the quadratic action for 
perturbations (''$\gamma$-crossing''). We describe several specific 
examples of bouncing cosmologies and models with Genesis epoch
which have neither ghosts nor gradient instabilities among the linearized 
perturbations about the homogeneous isotropic background during entire 
evolution.

\end{abstract}

\section{Introduction}
Cosmological scenarios with the bouncing or Genesis stage serve as 
possible extensions of the standard hot Big Bang theory.
In both of these scenarios,
space-time has vanishing 4D curvature at early times,
i.e., the Hubble parameter and its time derivatives take on small values.
The bouncing model implies that the Universe undergoes a contracting 
stage at early times, which terminates at some  moment of 
time (the bounce) and the Universe transits to the expansion epoch 
(see Refs.~\cite{bounce1,bounce2,bounce3} for reviews).
The Genesis scenario describes the accelerated expansion of the 
Universe from the asymptotically empty Minkowski space: the energy density
of exotic matter, which drives the evolution during the Genesis epoch, 
grows in time, and so does the expansion rate (the Hubble parameter);
at some stage, when the energy density and Hubble parameter have 
grown sufficiently large, the energy of the exotic matter gets transformed into 
heat, so that the Universe transits to the standard 
hot stage~\cite{gen1,gen2}.

The specific feature of both scenarios is the absence of initial
singularity, whose inevitable presence in the hot Big Bang theory
has not been overcome  even with the invention of
inflation~\cite{singularity_infl1,singularity_infl2}. 
In fact, non-singular cosmologies
with bounce or Genesis may be equally well
considered as complementary or alternative to
inflationary scenario~\cite{bounce3,bounpert1}.

One of the issues one should take care of when 
constructing bouncing models is Belinskii-Khalatnikov-Lifshitz 
(BKL) phenomenon~\cite{BKL}. It  may lead to strong
inhomogeneity and anisotropy of space at the end of  
contraction stage, which are unacceptable in a 
self-consistent cosmological model, 
see discussion in Refs.~\cite{Erickson2004} and~\cite{bounce1}. 
One of the possible solutions to the BKL problem within General Relativity
(GR) is to introduce  a matter component with a super stiff 
equation of state $p \geq \rho$ during the contraction stage, where
$\rho$ and $p$ are energy density and effective pressure, respectively.
One of the simple options involves a homogeneous
massless scalar field with the equation of state $p=\rho$  
dominating during the contraction
epoch; this option
follows from the results obtained by Khalatnikov and Kamenshchik
in Ref.~\cite{Khalatnikov:2003ph}.
Another approach to solving the BKL problem is realized,  for example, in
the ekpyrosis scenario~\cite{ekpy}. In any case, one of the viability
criteria for a bouncing model is the absence of the BKL behaviour 
during contraction.

An important property
of non-singular cosmologies with bounce or Genesis is the
necessity to introduce a
specific matter component, which, unless
one abandons GR or relies upon the 
3D spatial curvature, has to violate the Null Energy Condition (NEC), see, for instance, Ref.~\cite{RubakovNEC} for a review,
\be
\label{intro:NEC}
T_{\mu\nu} n^{\mu} n^{\nu} > 0,
\ee
where $T_{\mu\nu}$ is the energy-momentum tensor,
and $n^{\mu}$ is any null vector ($g_{\mu\nu} n^{\mu} n^{\nu} = 0$). 
In a general case, when gravity is modified, NEC is replaced with
the Null Convergence Condition (NCC)~\cite{NCC}
\be
\label{intro:NCC}
R_{\mu\nu} n^{\mu} n^{\nu} > 0,
\ee
where $R_{\mu\nu}$ is Ricci tensor.
The need to introduce the NEC-violating matter becomes evident 
upon considering the combination of Einstein equations for a spatially-flat,
homogeneous and isotropic Universe:
\be
\label{intro:Hubble_dot}
\dot{H} = - 4\pi G (\rho + p),
\ee
where $H$ denotes the Hubble parameter.
Indeed, by choosing $n^{\mu} = (1,a^{-1} q^i)$,
where ${\bf q}^2 = 1$, in the cosmological background described above 
the NEC~\eqref{intro:NEC} takes the form 
\be
\label{intro:NEC_cosmo}
p + \rho > 0.
\ee
If the NEC~\eqref{intro:NEC_cosmo} is satisfied, 
it follows from
eq.~\eqref{intro:Hubble_dot} that $\dot{H} < 0$, hence, 
if the Universe was contracting in the past, it would continue contraction
until it would reach singularity
\footnote{This reasoning does not apply to the case of the closed
  Universe, where the bounce is possible if the energy density and effective pressure grow slower than $a^{-2}$ during contraction~\cite{Sph_bounce}.}.
Likewise, NEC-violating matter  is necessary for the Universe starting off with Genesis: this scenario implies
growing Hubble parameter during the Genesis epoch which is forbidden 
by eqs.~\eqref{intro:Hubble_dot}, \eqref{intro:NEC_cosmo}.
Therefore, the cosmological solutions with bounce and
Genesis make it necessary to consider NEC-violating matter
(or NCC-violation in the case of modified gravity).
The latter feature makes bouncing and Genesis models
non-standard, since the majority of known types of matter
comply with the NEC/NCC while the attempts to violate 
these conditions often result in pathologies like ghosts,
gradient instabilities and tachyons among the linearized perturbations 
about the homogeneous isotropic background, see, for instance, 
Refs.~\cite{RubakovNEC,KobaRev} for reviews.

One of the possible ways to obtain the NEC/NCC-violation is to invoke 
the generalized Galileon theory~\cite{gal1,gal2,gal3,gal4}
or, equivalently, Horndeski theory~\cite{Horndeski}. Horndeski theories are 
the most general scalar-tensor theories of modified gravity with second 
derivative terms in the Lagrangian, whose presence, however, does not 
affect the order of differential equations of motion -- they are still second 
order in derivatives. Therefore, due to a specifically designed structure 
of the Lagrangian, Horndeski theories are free of Ostrogradsky ghosts and
 have 
$(2+1)$ dynamical degrees of freedom (DOF), two tensor modes 
and one scalar mode about the homogeneous isotropic background.
Quite recently an even more general class of scalar-tensor theories 
with second derivatives in the Lagrangian but without Ostrogradsky 
instabilities has been discovered -- the so-called
''degenerate higher order scalar-tensor theories'' (or DHOST 
theories)~\cite{Zuma,Gleyzes,Gleyzes2,DHOST2,DHOST3,DHOST4,DHOST5,DHOST6,DHOST7} 
and ``U-degenerate theories''~\cite{DeFelice:2018mkq}.
The important difference between these
generalizations and Horndeski theories is the fact that the former have 
third order equations of motion while propagating the same three DOF as 
Horndeski theories do. Moreover, there is a non-trivial relation between
some of the subclasses of DHOST theories 
and Horndeski theories via the invertible disformal 
transformation of metric~\cite{disformal,disf0,disf1,disf2}
\be
\label{intro:disformal}
g_{\mu\nu} \to \Omega^2(\pi, X) g_{\mu\nu} + \Gamma(\pi,X) \partial_{\mu}\pi\partial_{\nu}\pi,
\ee
where $\pi$ denotes a scalar field (Galileon field),
$X = g^{\mu\nu} \partial_{\mu}\pi\partial_{\nu}\pi$,
while $ \Omega^2(\pi, X)$ and $\Gamma(\pi,X)$ are arbitrary functions.
Interestingly, one of the first examples of DHOST theories from the
subclass called ''beyond Horndeski theories'' was discovered 
by applying the disformal transformation~\eqref{intro:disformal}
to a certain Lagrangian of Horndeski type~\cite{Zuma}.

Horndeski theories and their generalizations possess 
a specific feature of admitting healthy NEC/NCC-violating regimes. 
Here healthy means that violation of the NEC/NCC does not forbid the 
solution stability at the linearized level~\cite{KobaRev};
in what follows stability of a solution means that there are no ghost 
or gradient instabilities. 
The above feature makes this class of scalar-tensor theories interesting
from the viewpoint of
 construction of non-standard cosmologies 
like bounce and Genesis. And, indeed,  a 
significant number of bouncing scenarios were suggested
where the stage with the
NEC/NCC-violation was driven by the Galileon field of Horndeski 
type~\cite{bounG1,bounG2,bounG3,bounG4,bounG5,bounG6,bounG7,bounG8}.
These solutions were shown to be stable during some finite period of time, 
including the stage with the NEC/NCC-violation. 
The Universe with Genesis was also studied in
various subclasses of Horndeski 
theories~\cite{gen2,genG2,genG3,genG4,genG5,genG6,genG7,genG8}.
The issue of superluminal propagation of perturbation modes 
in the original Genesis model was specifically addressed
in Refs.~\cite{subgenG1,subgenG2}.

Further studies have shown, however, that there is
stability-related obstruction to
constructing complete non-singular
cosmological scenarios in Horndeski theories, i.e., the models whose
evolution can be 
followed from $t\to -\infty$ to $t\to + \infty$. 
Initially, a no-go theorem has been established for a cubic subclass
of Horndeski theories: it stated that there are no completely stable
solutions with bounce or Genesis since gradient instabilities and/or
ghosts inevitably arise sooner or later
among the perturbations about the homogeneous
isotropic background~\cite{LMR}. Similar property
has been found in the case when 
along with the Galileon field, 
there is an additional scalar field obeying the NEC~\cite{bounworm}.
These no-go theorems were further generalized to
Horndeski theories of the most general
form~\cite{Koba_nogo} as well as multi-Galileon
theories~\cite{Koba_nogo_multi}.
Hence, it has been shown on general grounds
that Horndeski theories are not suitable for constructing non-singular cosmological solutions that are stable during entire evolution (see also
Refs.~\cite{KobaRev,Mironov_nogo}).

The topic got new twist when the no-go argument and ways 
to circumvent it were analyzed within one of the subclasses of 
DHOST/U-degenerate 
theories dubbed beyond Horndeski theories or GLVP~\cite{Gleyzes,Gleyzes2}. 
Within  the effective field theory (EFT) approach, it was shown 
in Refs.~\cite{Cai:2016thi,Creminelli} that the beyond Horndeski terms
in the Lagrangian introduce significant changes to the stability condition,
which is a crucial ingredient of
the no-go theorem discussed above. This was
a strong indication that the no-go theorem might be evaded by going beyond 
Horndeski. The first explicit examples of stable non-singular 
cosmological solutions were suggested in Refs.~\cite{bouncegen,CaiBounce},
where a 
covariant approach was used instead of EFT.
The covariant formalism has the advantage of 
dealing directly with the Lagrangian of the theory, and, hence, 
enabling one to
check the solutions against the  equations of motion, which
is impossible in the EFT approach. As a result, Refs.~\cite{bouncegen,CaiBounce}
give explicit Lagrangians of beyond Horndeski type, which admit 
completely stable bouncing solutions; 
in Ref.~\cite{bouncegen} a complete
stable Genesis-like solution was constructed as well.

The solutions in Ref.~\cite{bouncegen} have a specific property that has to do
with the asymptotic behaviour of the theory as
$t \to \pm \infty$: at late times the beyond Horndeski theory transforms
into the theory of a conventional massless scalar field within GR, while
at early times ($t \to -\infty$) the Lagrangian does not simplify
and remains of beyond Horndeski type, which significantly differs from
GR + conventional scalar field. The simple form of the theory in
both asymptotics $t \to \pm \infty$ is
not an obligatory requirement, but it may be an advantage 
in the context of the further applications towards constructing
realistic models of the early Universe. 
As  discussed in Ref.~\cite{bouncegen} having the bounce or Genesis
with simple form of asymptotics is non-trivial because of 
the so-called $\gamma$-crossing phenomenon, which was
considered unacceptable at the time. We discuss in detail
the issue of $\gamma$-crossing in Sec.~\ref{sec:gamma}. There we 
stress that, in fact, there is nothing wrong with $\gamma$-crossing, as 
shown in Refs.~\cite{Ijjas,bounceI}. Once 
the healthy nature of $\gamma$-crossing was understood,
there were suggested 
completely stable bouncing and Genesis scenarios in beyond Horndeski 
theory, whose asymptotics at both
$t\to +\infty$ and $t\to -\infty$ are
described by GR with a massless scalar 
field~\cite{bounceI,genesisGR}. The additional advantage of the bouncing 
model of Ref.~\cite{bounceI} is the absence of the BKL phenomenon during
the contracting stage due to the
domination of the massless scalar field prior to the bounce.

To summarize,
today there are examples of completely stable cosmological solutions of 
bouncing and Genesis types within beyond Horndeski theory. However, one may
be puzzled by an apparent
contradiction: on the one hand, the no-go theorem valid 
in Horndeski subclass is evaded by going beyond Horndeski, but, on the 
other hand, Horndeski and beyond Horndeski subclasses are related by 
disformal transformation~\eqref{intro:disformal}.
Indeed, 
the disformal transformation~\eqref{intro:disformal}
is a mere field redefinition, which at a glance cannot affect the stability 
of a solution. The resolution of this apparent
paradox is that the
disformal transformation from the beyond Horndeski theory
with a completely stable solution to
the Horndeski theory with no-go theorem
turns out to be singular at a certain moment of time.
The latter result has been obtained within the
EFT approach in Ref.~\cite{Creminelli}. One of the purposes of this review is
to confirm this singular character of the disformal transformation 
in the covariant formalism utilised in 
Refs.~\cite{bouncegen,bounceI,genesisGR}, see Sec.~\ref{sec:disformal} 
for details.

This brief review has the following structure. In Sec.~\ref{sec:review}
we revisit the construction and stability analysis of the 
cosmological models with bounce and Genesis in beyond Horndeski theory.
In particular,  Sec.~\ref{sec:no-go} discusses the no-go theorem
for Horndeski theory as well as ways to circumvent it;  
Sec.~\ref{sec:gamma} considers the nature of $\gamma$-crossing
and specifies its role in  bouncing and Genesis solutions. 
In Sec.~\ref{sec:examples} we briefly describe the reconstruction 
procedure for obtaining completely stable solutions with 
bounce and Genesis 
in beyond Horndeski
theories and revisit explicit examples of these models suggested 
in Refs.~\cite{bouncegen,bounceI,genesisGR}. In Sec.~\ref{sec:disformal}
we discuss the relation of Horndeski theories with their extensions
via disformal transformation and show, in the
covariant formalism, that 
beyond Horndeski theories with completely stable non-singular solutions
are related to  Horndeski theories via field redefinition 
which is inevitably singular. We conclude in Sec.~\ref{sec:conclusion}. 


\section{Stability of non-singular cosmological scenarios in beyond
Horndeski theory
}
\label{sec:review}
This section reviews the existing results on the construction of
cosmological solutions and their stability analysis
in Horndeski theories and beyond.
In what follows we make use of notations introduced 
in Ref.~\cite{Kobayashi}, which have been
later adopted in Refs.~\cite{bouncegen,bounceI,genesisGR}.


\subsection{Lagrangian and stability conditions}
The general form of beyond Horndeski Lagrangian reads
(metric signature $(+,-,-,-)$):
\bea
  \label{lagrangianBH}
   &&
    \phantom{\mathcal{L}_2=F(\pi,X)\quad}
  S = \int d^4x \sqrt{-g}\Big( \mathcal{L}_2 + \mathcal{L}_3 + \mathcal{L}_4 + \mathcal{L}_5
   + \mathcal{L_{BH}}\Big),\\
  \label{L2H}
  &&\mathcal{L}_2=F(\pi,X),\\
  \label{L3H}
  &&\mathcal{L}_3=K(\pi,X)\Box\pi,\\
  \label{L4H}
  &&\mathcal{L}_4=-G_4(\pi,X)R+2G_{4X}(\pi,X)\left[\left(\Box\pi\right)^2-\pi_{;\mu\nu}\pi^{;\mu\nu}\right],\\
  \label{L5H}
  &&\mathcal{L}_5=G_5(\pi,X)G^{\mu\nu}\pi_{;\mu\nu}+\frac{1}{3}G_{5X}\left[\left(\Box\pi\right)^3-3\Box\pi\pi_{;\mu\nu}\pi^{;\mu\nu}+2\pi_{;\mu\nu}\pi^{;\mu\rho}\pi_{;\rho}^{\;\;\nu}\right],
  \\
  \label{LBH}
  &&\mathcal{L_{BH}}=F_4(\pi,X)\epsilon^{\mu\nu\rho}_{\quad\;\sigma}\epsilon^{\mu'\nu'\rho'\sigma}\pi_{,\mu}\pi_{,\mu'}\pi_{;\nu\nu'}\pi_{;\rho\rho'}+
  \\\nonumber&&\qquad+F_5(\pi{},X)\epsilon^{\mu\nu\rho\sigma}\epsilon^{\mu'\nu'\rho'\sigma'}\pi_{,\mu}\pi_{,\mu'}\pi_{;\nu\nu'}\pi_{;\rho\rho'}\pi_{;\sigma\sigma'},
\eea
where $\pi$ is the scalar (Galileon) field,
$X=g^{\mu\nu}\pi_{,\mu}\pi_{,\nu}$, $\pi_{,\mu}=\partial_\mu\pi$,
$\pi_{;\mu\nu}=\triangledown_\nu\triangledown_\mu\pi$,
$\Box\pi = g^{\mu\nu}\triangledown_\nu\triangledown_\mu\pi$,
$G_{4X}=\partial G_4/\partial X$, etc.;
$R$ in eq.~\eqref{L4H} and $G^{\mu\nu}$ in eq.~\eqref{L5H} denote
Ricci scalar and Einstein tensor, respectively.
The terms~\eqref{L2H}~--~\eqref{L5H} describe Horndeski theory
and
involve 4 independent functions $F(\pi,X)$, $K(\pi,X)$,
$G_4(\pi,X)$ and $G_5(\pi,X)$.
The functions $F_4(\pi,X)$ and $F_5(\pi,X)$ in eq.~\eqref{LBH} are characteristic of beyond Horndeski theory.
Let us note that the action~\eqref{lagrangianBH} already contains the
gravitational part (see eqs.~\eqref{L4H} and~\eqref{L5H}):
the Einstein--Hilbert action is restored by setting
$G_4(\pi,X) = 1/2\kappa$ and $G_5(\pi,X) = 0$, where $\kappa = 8\pi G$ 
and $G$ is the gravitational constant. 
The Lagrangian for cubic Horndeski theory, which was mentioned above and extensively used in recent works, reads 
\be
\label{cubicH}
\mathcal{L}_{{cub}} = - \frac{1}{2\kappa} R + \mathcal{L}_2+\mathcal{L}_3.
\ee
By adding terms \eqref{L4H} and~\eqref{L5H} to the
Lagrangian~\eqref{cubicH},
one obtains quartic and quintic Horndeski theories, respectively.

Clearly, the Lagrangian of the general Horndeski theory, 
i.e., the theory with  $F_4=F_5=0$,
contains the second
derivatives of both the Galileon field $\pi$ and metrics.
Generally, these second derivatives cannot be removed by
integration by parts. Nevertheless, all  field equations
are differential equations of the second order at most. Beyond Horndeski
theories with $F_4\neq 0$ and/or $F_5 \neq 0$ 
do not have this property;
however, as we mentioned in Introduction, these theories 
propagate the same number of DOF as the general Horndeski theories,
i.e., two tensor and one scalar modes. The same is true for even
more general classes of DHOST and U-degenerate theories, whose Lagrangians
are not given in this review, see Refs.~\cite{DHOST7,DeFelice:2018mkq}:
it is sufficient for our purposes to consider theories with the 
Lagrangian~\eqref{lagrangianBH}. Moreover, to make the
formulas that follow more concise we take
\[
G_5 = 0 \; , \;\;\;\;\;\; F_5 = 0 \;.
\]
There is nothing fundamentally new for our studies in the general
case with $G_5 \neq 0$, $F_5 \neq 0$, while the formulas become
cumbersome.

We consider the cosmological models described by  spatially flat 
Friedmann--Lema\^itre--Robertson--Walker (FLRW) metric:
\begin{equation}
\label{FLRW}
\mathrm{d}s^2 = \mathrm{d}t^2 - a^2(t)\delta_{ij}\mathrm{d}x^i\mathrm{d}x^j.
\end{equation}
In our setup, the background Galileon field is homogeneous, 
$\pi = \pi (t)$.
In this case the independent field equations for the theory
with the action~\eqref{lagrangianBH} read:
\bea
\label{dg00}
\delta g^{00}:\quad
&&F-2F_XX-6HK_XX\dot{\pi}+K_{\pi}X+6H^2G_4+6HG_{4\pi}\dot{\pi}
\\\nonumber&&-24H^2X(G_{4X}+G_{4XX}X)+12HG_{4\pi X}X\dot{\pi}
-6H^2X^2(5F_4+2F_{4X}X) = 0 \; ,\\
\label{dgii}
\delta g^{ii}:\quad
&&F-X(2K_X\ddot{\pi}+K_\pi)+2(3H^2+2\dot{H})G_4-12H^2G_{4X}X
\\\nonumber&&-8\dot{H}G_{4X}X-8HG_{4X}\ddot{\pi}\dot{\pi}-16HG_{4XX}X\ddot{\pi}\dot{\pi}
\\\nonumber&&+2(\ddot{\pi}+2H\dot{\pi})G_{4\pi}+4XG_{4\pi X}(\ddot{\pi}-2H\dot{\pi})+2XG_{4\pi\pi}
\\ \nonumber&& -2F_4X(3H^2X+2\dot{H}X+8H\ddot{\pi}\dot{\pi})-8HF_{4X}X^2\ddot{\pi}\dot{\pi}
-4HF_{4\pi}X^2\dot{\pi} = 0
\eea
where $H=\dot{a}/a$ is the Hubble parameter. The field equation
obtained by varying the action~\eqref{lagrangianBH} 
over $\pi$ is a linear combination
of eqs.~\eqref{dg00},~\eqref{dgii} and their derivatives.

The central issue for cosmological models is their
stability under linearized inhomogeneous perturbations.
In the linearized theory we study both metric perturbations 
and scalar field perturbations $\pi$. Let us introduce the
following notations for the metric components, which include
both background and linearized perturbations:
\begin{equation}
\label{linearized_metric}
g_{00}=1+2\alpha,\quad g_{0i}=-\partial_i\beta,\quad g_{ij}=-a^2\left(2\zeta\delta_{ij}
+h_{ij}^T\right),
\end{equation}
where $\alpha$, $\beta$ and $\zeta$ are scalar perturbations,
$h_{ij}^T$ stand for tensor modes, which are traceless
($h_{ii}^T = 0$) and transverse
($\partial_i h_{ij}^T = 0$).
Note that there are no non-vanishing vector perturbations
in the scalar-tensor theories in question, and that we have partly
used gauge freedom in the
parametrization~\eqref{linearized_metric}.
The perturbation about the homogeneous background Galileon 
field $\pi_c$ is denoted by $\chi$:
\be
\label{chi}
\pi \rightarrow \pi_c(t) + \chi(t,r).
\ee
Generally, the linearized theory is invariant under infinitesimal
coordinate transformations of the following form:
\be
\label{general_transform}
x^{\mu} \rightarrow x^{\mu} + \xi^{\mu},
\ee
where $\xi^{\mu}$ are infinitesimal parameters.
Part of this gauge freedom has been already used 
in eq.~\eqref{linearized_metric}.
The residual gauge freedom is parametrized by the gauge function $\xi^0$. 
In terms of the parametrization  
\eqref{linearized_metric} and~\eqref{chi},
the transformation law 
for the scalar modes reads:
\begin{equation}
\label{gauge_transform}
\chi \to \chi + \xi^0\dot{\pi},
\quad \alpha \to \alpha + \dot{\xi^0},
\quad \beta \to \beta - \xi^0,
\quad \zeta \to \zeta + \xi^0\dfrac{\dot{a}}{a}.
\end{equation}
We fix the residual gauge freedom by setting
$\chi=0$ (unitary gauge),
so that the only non-trivial
modes in the scalar sector are $\alpha$, $\beta$ and $\zeta$.
Then the quadratic action for perturbations in
theory~\eqref{lagrangianBH} has the following form:
\begin{equation}
\label{quadraction1}
\begin{aligned}
S=\int\mathrm{d}t\mathrm{d}^3x a^3
\Bigg[\left(\dfrac{\mathcal{{G}_T}}{8}\left(\dot{h}^T_{ij}\right)^2
-\dfrac{\mathcal{F_T}}{8a^2}\left(\partial_k h_{ij}^T\right)^2\right)
+\Big(-3\mathcal{{G}_T}\dot{\zeta}^2
+\mathcal{F_T}\dfrac{(\triangledown\zeta)^2}{a^2}\\
-2(\mathcal{{G}_T}+\mathcal{D}\dot{\pi})\alpha\dfrac{\triangle\zeta}{a^2}
+2\mathcal{{G}_T}\dot{\zeta}\dfrac{\triangle\beta}{a^2}
+6\Theta\alpha\dot{\zeta}
-2\Theta\alpha\dfrac{\triangle\beta}{a^2}
+\Sigma\alpha^2\Big)\Bigg],
\end{aligned}
\end{equation}
where
$(\triangledown\zeta)^2 = \delta^{ij} \partial_i \zeta \partial_j \zeta$,
$\triangle = \delta^{ij} \partial_i \partial_j$, and
coefficients $\mathcal{G_T}$, $\mathcal{D}$, $\mathcal{F_T}$,
$\Theta$ and $\Sigma$ are expressed in terms of the Lagrangian functions:
\bea
\label{eq:GT_coeff_setup}
&&\mathcal{G_T}=2G_4-4G_{4X}X
- 2F_4X^2
,\\
\label{eq:D_coeff_setup}
&&\mathcal{D}=2F_4X\dot{\pi}
,\\
\label{eq:FT_coeff_setup}
&&\mathcal{F_T}=2G_4 
,\\
\label{eq:Theta_coeff_setup}
&&\Theta=-K_XX\dot{\pi}+2G_4H-8HG_{4X}X-8HG_{4XX}X^2+G_{4\pi}\dot{\pi}+2G_{4\pi X}X\dot{\pi}\\
\nonumber&&-10HF_4X^2-4HF_{4X}X^3
,\\
\label{eq:Sigma_coeff_setup}
&&\Sigma=F_XX+2F_{XX}X^2+12HK_XX\dot{\pi}+6HK_{XX}X^2\dot{\pi}-K_{\pi}X-K_{\pi X}X^2\\
\nonumber&&-6H^2G_4+42H^2G_{4X}X+96H^2G_{4XX}X^2+24H^2G_{4XXX}X^3
-6HG_{4\pi}\dot{\pi} \\
\nonumber&& -30HG_{4\pi X}X\dot{\pi}-12HG_{4\pi XX}X^2\dot{\pi} 
+90H^2F_4X^2+78H^2F_{4X}X^3+12H^2F_{4XX}X^4
\eea
We note that fixing the gauge directly in the 
quadratic action~\eqref{quadraction1} (rather than in the field equations)
is legitimate 
since
the Galileon field equation follows from eqs.~\eqref{dg00},~\eqref{dgii}
(see Ref.~\cite{Kobayashi} for discussion and 
Ref.~\cite{DeFelice:2018mkq} for details).

Due to the structure of the quadratic action~\eqref{quadraction1}, 
both $\alpha$ and $\beta$ are non-dynamical. Varying the 
action~\eqref{quadraction1} with respect to $\alpha$
and $\beta$, one obtains two constraint equations:
\bea
\label{eq:constr_beta}
\dfrac{\triangle\beta}{a^2} &=& \dfrac{1}{\Theta}\left(3\Theta\dot{\zeta}
-(\mathcal{{G}_T}+\mathcal{D}\dot{\pi})\dfrac{\triangle\zeta}{a^2}+\Sigma\alpha\right),\\
\label{eq:constr_alpha}
\alpha &=& \dfrac{\mathcal{{G}_T}\dot{\zeta}}{\Theta}.
\eea
By utilizing the constraints~\eqref{eq:constr_beta}
and~\eqref{eq:constr_alpha}, one recasts
the action~\eqref{quadraction1} in terms of dynamical DOF only:
\begin{equation}
\label{eq:unconstrained_action}
S=\int\mathrm{d}t\mathrm{d}^3x \;a^3
\left[\dfrac{\mathcal{{G}_T}}{8}\left(\dot{h}^T_{ij}\right)^2
-\dfrac{\mathcal{F_T}}{8a^2}\left(\partial_k h_{ij}^T\right)^2
+\mathcal{G_S}\dot{\zeta}^2
-\mathcal{F_S}\dfrac{(\triangledown\zeta)^2}{a^2}\right],\\
\end{equation}
where the following notations are introduced:
\bea
\label{eq:GS_setup}
&&\mathcal{G_S}=\dfrac{\Sigma\mathcal{{G}_T}^2}{\Theta^2}+3\mathcal{{G}_T},\\
\label{eq:FS_setup}
&&\mathcal{F_S}=\dfrac{1}{a}\dfrac{\mathrm{d}\xi}{\mathrm{d}t}-\mathcal{F_T},\\
\label{eq:xi_func_setup}
&&\xi
=\dfrac{a\left(\mathcal{{G}_T}+\mathcal{D}\dot{\pi}\right)\mathcal{{G}_T}}{\Theta}.
\eea
The quadratic action~\eqref{eq:unconstrained_action}
describes one scalar ($\zeta$) and two tensor ($h^T_{ij}$) DOF.
The propagation speed squared of
the scalar and tensor perturbations reads,
respectively:
\be
c_\mathcal{T}^2=\dfrac{\mathcal{F_T}}{\mathcal{{G}_T}},\qquad c_\mathcal{S}^2=\dfrac{\mathcal{F_S}}{\mathcal{G_S}}.
\ee

Let us comment on the main
types of instabilities, 
which can possibly
arise in the quadratic action~\eqref{eq:unconstrained_action}. 
In the case of  homogeneous and isotropic
background, the coefficients $\mathcal{G_T}$, $\mathcal{F_T}$, 
$\mathcal{G_S}$ and $\mathcal{F_S}$ are functions of time.
The most dangerous instabilities are those arising
in the high energy regime, i.e., when the characteristic scales of temporal
and spatial variations of $\zeta$ and $h^T_{ij}$ are considerably
smaller than that of the homogeneous background.
These are the instabilities that we consider in this review.
In the high energy
approximation, the coefficients  
$\mathcal{G_{S,T}}$ and $\mathcal{F_{S,T}}$ can be treated as
time-independent at relevant time intervals.
Then the following situations are possible 
(the notations $\mathcal{G_{S,T}}$, $\mathcal{F_{S,T}}$ refer
to pairs of coefficients $\mathcal{G_{S}}$, $\mathcal{F_{S}}$ or 
$\mathcal{G_{T}}$, $\mathcal{F_{T}}$):
\begin{itemize}
\item[(1)] Gradient instabilities (exponential growth 
of perturbations):
\be
\nonumber
\quad \mathcal{G_{S,T}} >0, \quad \mathcal{F_{S,T}} < 0, \quad \mbox{or}  \quad \mathcal{G_{S,T}} < 0, \quad \mathcal{F_{S,T}} > 0.
\ee
\item[(2)] Ghosts (catastrophic instability of
  vacuum state, see Ref.~\cite{RubakovNEC} for discussion):
\be
\quad\qquad \mathcal{{G_{S,T}}} < 0, \quad \mathcal{{F_{S,T}}} < 0.
\ee
\item[(3)] Stable solution:
\be
\label{eq:healthy}
\quad \mathcal{{G_{S,T}}} >0, \quad \mathcal{{F_{S,T}}} >0.
\ee
\end{itemize}
Let us note that due to the form of the action~\eqref{eq:unconstrained_action}
in the unitary gauge, the tachyonic instabilities do
not develop in the system.

Hence, according to eq.~\eqref{eq:healthy}, the absence of ghost and
gradient instabilities about a homogeneous background solution
implies the following restrictions on the coefficients 
in the quadratic action~\eqref{eq:unconstrained_action}: 
\begin{equation}
\label{eq:stability_cond}
\mathcal{{G}_T} \geq \mathcal{F_T}> \eps >0,\quad \mathcal{G_S}\geq\mathcal{F_S} >\eps> 0 \; .
\end{equation}
Hereafter $\eps$ denotes a positive constant,
whose actual value is irrelevant
for our reasoning, so it may be different  in different formulas below.
This constant is introduced in eq.~\eqref{eq:stability_cond}
to avoid the situations when $\mathcal{G_{S,T}},\mathcal{F_{S,T}} \to 0$,
which, at least naively, corresponds to  strong coupling
regime~\footnote{We do not consider the special case of  
ghost condensate~\cite{GHC}.}.
The inequalities~\eqref{eq:stability_cond} also ensure that
both scalar and tensor perturbations propagate at
the speed of light at most.

As we alluded to above, it was shown in Refs.~\cite{LMR,Koba_nogo}
that it is impossible to satisfy 
the constraints~\eqref{eq:stability_cond} over the entire evolution
in Horndeski theories with $F_4(\pi,X) = F_5(\pi,X) = 0$
(see eq.~\eqref{LBH}). This is precisely the no-go theorem
which states that in the general Horndeski theory
there are no completely stable bouncing and Genesis cosmologies. 
We discuss this no-go theorem in the next subsection
in order to clarify how the stability conditions
for cosmological solutions get modified in the presence of 
terms with $F_4(\pi,X)$
(and $F_5(\pi,X)$) in the Lagrangian.


\subsection{No-go theorem in Horndeski theory}
\label{sec:no-go}
The no-go theorem in Horndeski theory is obtained by the stability 
analysis of cosmological scenarios, under the assumption that the
scale factor $a$ is
bounded from below by a positive constant, which ensures the 
geodesic completeness. The theorem is based 
on the requirement of absence of gradient instabilities
(see eq.~\eqref{eq:FS_setup}):
\be
\label{nogo}
\dfrac{\mathrm{d}\xi}{\mathrm{d}t} =
a \left(\mathcal{F_S} +\mathcal{F_T}\right) > \eps > 0.
\ee
According to eq.~\eqref{nogo}, the  coefficient 
\be
\label{eq:xi_H}
\xi=\dfrac{a\mathcal{G_T}^2}{\Theta}
\ee
has to be a monotonously growing function of time.
Note that the definition of $\xi$ in eq.~\eqref{eq:xi_H} 
is valid only in Horndeski theory
where $\mathcal{D} = 0$ (cf. eq.~\eqref{eq:xi_func_setup}).
It follows from the constraint~\eqref{nogo}, which must hold
at any moment of time, and eq.~\eqref{eq:stability_cond}
that
$\xi \to -\infty$ as $t \to -\infty$ and $\xi \to +\infty$ as
 $t \to +\infty$,
and, hence, $\xi$ necessarily crosses zero at some moment(s) of time.
The latter fact is true irrespectively of whether or not the
coefficient $\Theta$ vanishes at some moment(s) of time.
Let us note that $\xi$ behaves as described above in beyond 
Horndeski theories as well (i.e., when $\mathcal{D} \neq 0$ and $\xi$
is defined by eq.~\eqref{eq:xi_func_setup}),
since the condition~\eqref{nogo} holds for both 
Horndeski and beyond Horndeski subclasses.
However, it follows from the definition~\eqref{eq:xi_H} that
in Horndeski theory, $\xi$ cannot behave in the way it is
supposed to: since $a > 0$ and $\mathcal{G_T}>\eps>0$, the only way 
$\xi$ can cross zero is when $\Theta \to \infty$, which in turn
corresponds to a singularity in the classical solution.
Thus, there are no completely stable bouncing and Genesis
models in Horndeski theory. This result still holds when
$G_5 \neq 0$ (but $F_4=F_5 =0$).

A comment is in order on attempts to evade the no-go theorem 
within the general Horndeski subclass~\cite{Koba_nogo,bounG8}.
One of the possible options is to make zero-crossing
of $\xi$ happen due to simultaneously vanishing $\Theta$ 
and $\mathcal{G_T}$, i.e., 
$\Theta(t_*) = 0$ and $\mathcal{G_T}(t_*)=0$, which
violates the conditions~\eqref{eq:stability_cond}.
This option not only implies fine-tuning, but  also faces the problem
of strong coupling in the tensor sector 
(see eq.~\eqref{eq:unconstrained_action}).
Another way to evade the no-go theorem  is to partly give up
the restrictions on the asymptotic behaviour of the theory, 
i.e., allow $\mathcal{F_{S,T}} \to 0$ as $t \to-\infty$ and/or 
$t \to+\infty$. This case is potentially problematic because of
naive strong coupling in the asymptotic past and/or asymptotic future.

The situation in beyond Horndeski theory is fundamentally different:
the definition of $\xi$, eq.~\eqref{eq:xi_func_setup}, 
involves $\mathcal{D} \neq 0$ due to the function
$F_4(\pi,X)$ (and $F_5(\pi,X)$).
While the coefficient $\mathcal{{G}_T}$ is still responsible for
stability in the tensor sector and has to be always positive,
the combination $(\mathcal{G_T} +\mathcal{D}\dot{\pi})$ is 
unconstrained and can take any values, including zero and negative ones.
It is due to this coefficient $\mathcal{D}$ that 
$\xi$ can monotonously grow and  cross zero at some moment
of time; the no-go theorem no longer holds.
Therefore, the form of the stability 
conditions~\eqref{eq:stability_cond} in beyond Horndeski theories 
points towards the opportunity to construct bouncing and Genesis solutions
free of ghost and gradient instabilities
during entire evolution.


\subsection{$\gamma$-crossing}
\label{sec:gamma}

Before we move on to explicit examples of bouncing and Genesis
solutions in beyond Horndeski theory, let us discuss the possible 
behaviour of the coefficient $\Theta$ in eq.~\eqref{eq:xi_func_setup}
and, in particular, the so-called $\gamma$-crossing phenomenon, meaning
$\Theta = 0$
(in Refs.~\cite{bounG8,Ijjas,Vikman}, 
where the issue was originally addressed, the notations differ and 
the coefficient $\Theta$ is denoted by $\gamma$, which explains the 
terminology).
In Ref.~\cite{Vikman} it was shown that $\gamma$-crossing occurs at
the change of the branch of the solution to
eq.~\eqref{dg00}, 
considered as quadratic 
equation for the Hubble parameter.
It was mentioned above that $\gamma$-crossing does not help
circumvent the no-go theorem in Horndeski theory. 
Instead, this phenomenon plays a crucial role in
determining the asymptotic behaviour of a beyond Horndeski
theory as $t \to \pm \infty$. Namely,
if we require that the beyond Horndeski
theory which admits
the non-singular cosmological solution in question,
tends to GR with, say, a conventional massless scalar field in both
asymptotic past and
asymptotic future, the function $\Theta (t)$ must cross
zero at some moment $t_* \in (-\infty,+\infty)$.
Indeed, these asymptotics imply that
$F_4 \to 0$ as $t \to \pm \infty$ and, consequently,	
$\mathcal{D} \to 0$  as $t \to \pm \infty$ 
(see eq.~\eqref{eq:D_coeff_setup}).
As we argued above, 
$\xi < 0$ as $t\to -\infty$ and $\xi > 0$ as $t\to +\infty$;
together with $\mathcal{D} \to 0$  as $t \to \pm \infty$ this
means that 
$\Theta < 0$ as $t\to -\infty$ and
$\Theta > 0$ as $t\to +\infty$, which proves that the 
coefficient $\Theta$ vanishes at some finite moment of time $t_*$.

However, the expressions for
$\mathcal{{G}_S}$ and $\mathcal{{F}_S}$
in eqs.~\eqref{eq:GS_setup},~\eqref{eq:FS_setup} show singular
behaviour of both coefficients at the moment of $\gamma$-crossing,
$\Theta = 0$.
At a glance this appears unacceptable. This was the reason
for requiring that $\Theta$ is always
positive when constructing one of the first completely stable
bouncing solutions in Ref.~\cite{bouncegen}.
In full accordance with the discussion above, the forbidden
$\gamma$-crossing made it impossible to design a bouncing model
whose asymptotics as $t \to \pm \infty$ are both
described by GR, so gravity in the solution of Ref.~\cite{bouncegen}
significantly differs from GR in the asymptotic past.
But, interestingly, 
the same expressions~\eqref{eq:GS_setup}, \eqref{eq:FS_setup}
indicate that the dispersion relation 
$c_\mathcal{S}^2 = \mathcal{F_S}/\mathcal{G_S}$ remains
finite at $\gamma$-crossing, which in turn suggests that
the situation is in fact not pathological. 
And, indeed, it was shown in Ref.~\cite{Ijjas} that the equations
for perturbations in the Newtonian gauge
do not exhibit singularities when $\Theta =0$.
Later similar calculations have been carried out in the 
unitary gauge~\cite{bounceI}, and it has been found that
the solution for the scalar DOF $\zeta$ is regular  at 
$\gamma$-crossing. Therefore, it has been proven that $\gamma$-crossing
is acceptable. The latter observation
made it possible to construct 
 cosmological solutions with bounce and Genesis, whose
 asymptotics are simple as $t \to \pm \infty$ so that the theory
 reduces there 
to GR + a conventional massless scalar 
field~\cite{bounceI,genesisGR}.

In the following subsection we review the explicit examples of
models with completely stable bounce and Genesis in beyond Horndeski
theories~\cite{bouncegen,bounceI,genesisGR}.
We highlight the way the no-go theorem is evaded in each of these 
solutions and describe their specific features.


\subsection{Completely stable models with bounce and Genesis: examples}
\label{sec:examples}
We do not go into details of the construction  of
solutions,
which are described in Refs.~\cite{bouncegen,bounceI,genesisGR}.
Instead, we  focus on  the main ideas and results.

One way to design the models in question is to employ the reconstruction
method, which was extensively used in the previous works, e.g.,
in Refs.~\cite{bounG7,LMR}\footnote{ Let us also mention
the discussion in Ref.~\cite{Vikman} of the solutions  of Ref.~\cite{bounG7}.}.
The general strategy is to find the Lagrangian functions
$F$, $K$, $G_4$, $F_4$ in eqs.~\eqref{L2H} -- \eqref{LBH}
such that the theory with the Lagrangian~\eqref{lagrangianBH} 
admits the desired solution (we still take $G_5=F_5=0$).
In the first place, by making use of  field redefinition 
it is always possible to choose the linear Galileon background
\be
\label{pi}
\pi_c(t) = t \; .
\ee
Then $X_c = g^{\mu \nu}_c \partial_\mu \pi_c \partial_\nu \pi_c = 1$.
Now, the field equations~\eqref{dg00}, \eqref{dgii} and stability
conditions~\eqref{eq:stability_cond} involve functions of 
time $F(\pi_c, X_c) = F(t,1)$, $F_X (\pi_c ,X_c) = (\partial F /\partial X)
(t,1) \equiv F_X(t,1)$, etc., which are independent of each other
(while, for instance, the function $G_{4, \pi} $
equals $\dot{G}_4 (t, 1)$).
The aim is to select these functions for the explicitly 
specified Hubble parameter $H(t)$. This selection 
should satisfy the following requirements: 
(i) the field equations~\eqref{dg00}, \eqref{dgii} should hold; 
(ii) the solution must be stable, i.e., the stability 
conditions~\eqref{eq:stability_cond} with the coefficients~\eqref{eq:GT_coeff_setup}-- \eqref{eq:Sigma_coeff_setup}
should be satisfied. Clearly,
these requirements do not uniquely determine all  functions
$F(t,1)$, $F_X(t,1) , \dots, F_{4 XX}(t,1)$ entering 
eqs.\eqref{dg00},~\eqref{dgii},~\eqref{eq:stability_cond}:
there are only two equations, while the stability 
conditions~\eqref{eq:stability_cond} are 
inequalities rather than equations. Hence, the reconstruction we discuss
has high degree of arbitrariness,
and some of the functions are chosen
on simplicity basis.

We also impose the constraint on the  asymptotic behaviour
of the theory as $t \to\pm\infty$, which is not an obligatory requirement
but rather a matter of choice. Namely,
below we arrange the solutions in such a way that
in the asymptotic future ({\itshape and} in the asymptotic past 
in cases~\ref{subsec:two} and~\ref{subsec:three}) the beyond Horndeski
theory tends to GR with a conventional
massless scalar field.
Let us recall that the massless scalar field minimally coupled to 
gravity has the equation of state $p=\rho$, so that
the spatially flat solution in GR has the following form:
\[
a(t) \propto |t|^{1/3} \; , \;\;\;\;\;\; H(t) = \frac{1}{3t} \; ,
\]
while the canonical scalar field behaves as
$\phi_c (t) = \pm \sqrt{2/3} \ln |t|$. In view of eq.~\eqref{pi},
it is related to $\pi$ by
\[
\pi = \mbox{exp} \left( \sqrt{\frac{3}{2}} \phi \right)\; .
\]
This
must hold in the corresponding asymptotic. Here and in what follows
we set
\[
\kappa = 8\pi G =1 \; .
\]


\subsubsection {Cosmological bounce with an exotic contraction stage}

One of the first examples of a completely stable bouncing solution
with an explicitly constructed beyond Horndeski Lagrangian
is given in Ref.~\cite{bouncegen}.
This solution has a characteristic feature of forbidden $\gamma$-crossing,
which, as we discussed above, makes it impossible to have a
completely stable solution with GR asymptotics both as $t \to +\infty$
and  $t\to - \infty$.
So, we set $\Theta > 0$ at all times  and require that the theory reduces to
 GR + massless scalar field 
 only in the future asymptotics $t\to +\infty$.
 
 Within the reconstruction approach, the scale factor and hence
 the Hubble parameter are chosen at one's will. In this model,
 a simple choice is made:
\begin{equation}
\label{eq:hubble_bouncegen}
H(t) = \dfrac{t}{3(\tau^2 + t^2)}, \qquad 
a(t) = (\tau^2 + t^2)^\frac{1}{6},
\end{equation}
so that the bounce occurs at $t = 0$; $\tau$ is a parameter 
which defines the duration of the bouncing epoch (in what follows
we set $\tau=10$ for definiteness), while the asymptotic behaviour
$H(t)|_{t \to +\infty} \to (3t)^{-1}$ agrees with the 
required property of the theory as $t \to +\infty$.
According to the reconstruction procedure, we choose  part of the
Lagrangian
functions in such a way that  the stability 
conditions~\eqref{eq:stability_cond} and asymptotic constraints are satisfied, 
and then find the rest of functions from
the background equations of motion with
$H(t)$ given by eq.~\eqref{eq:hubble_bouncegen}.

\begin{figure}[h!]
\begin{minipage}[h!]{0.49\linewidth}
\center{\includegraphics[width=1\linewidth]{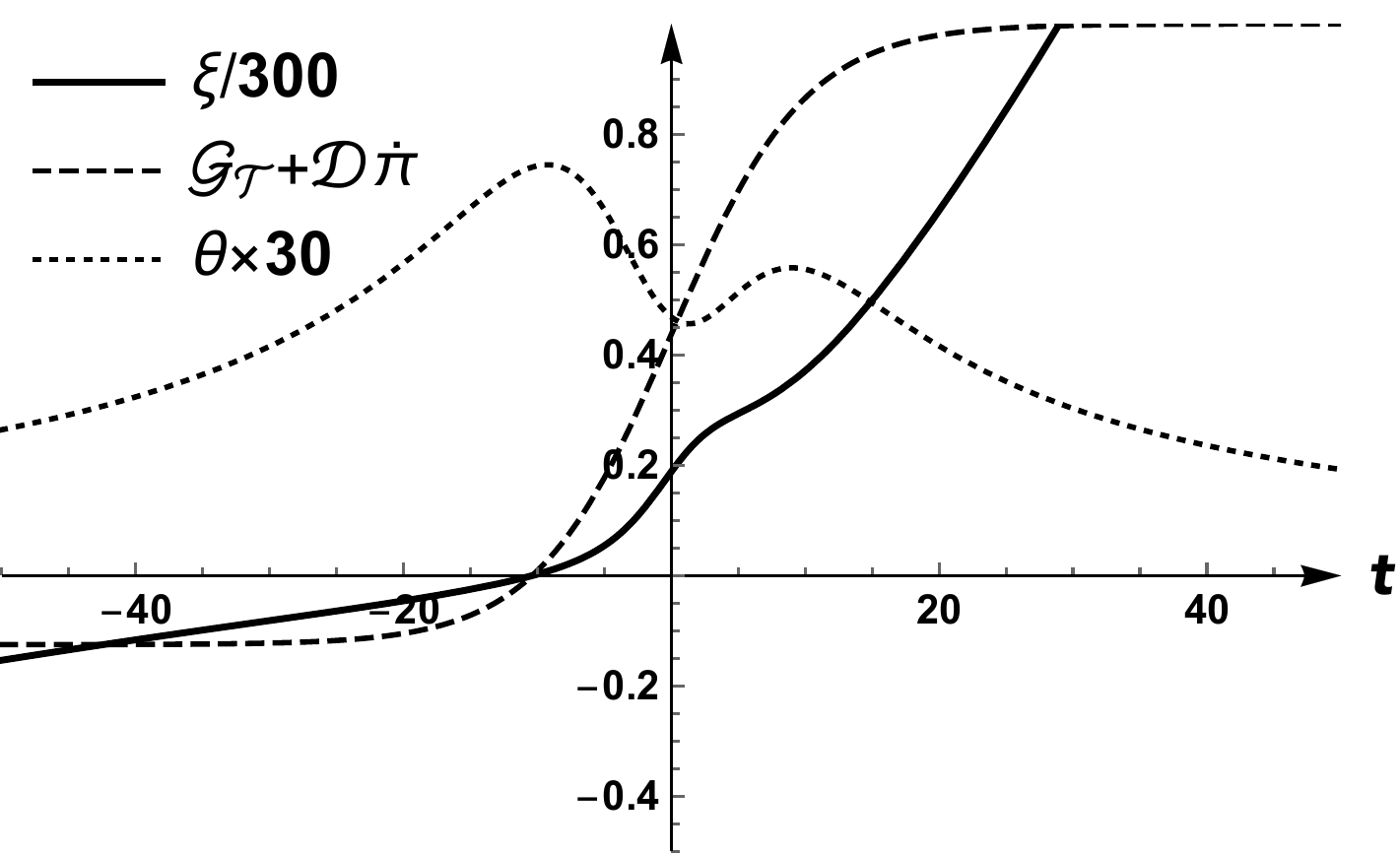}} (a) \\
\end{minipage}
\hfill
\begin{minipage}[h!]{0.49\linewidth}
\center{\includegraphics[width=1\linewidth]{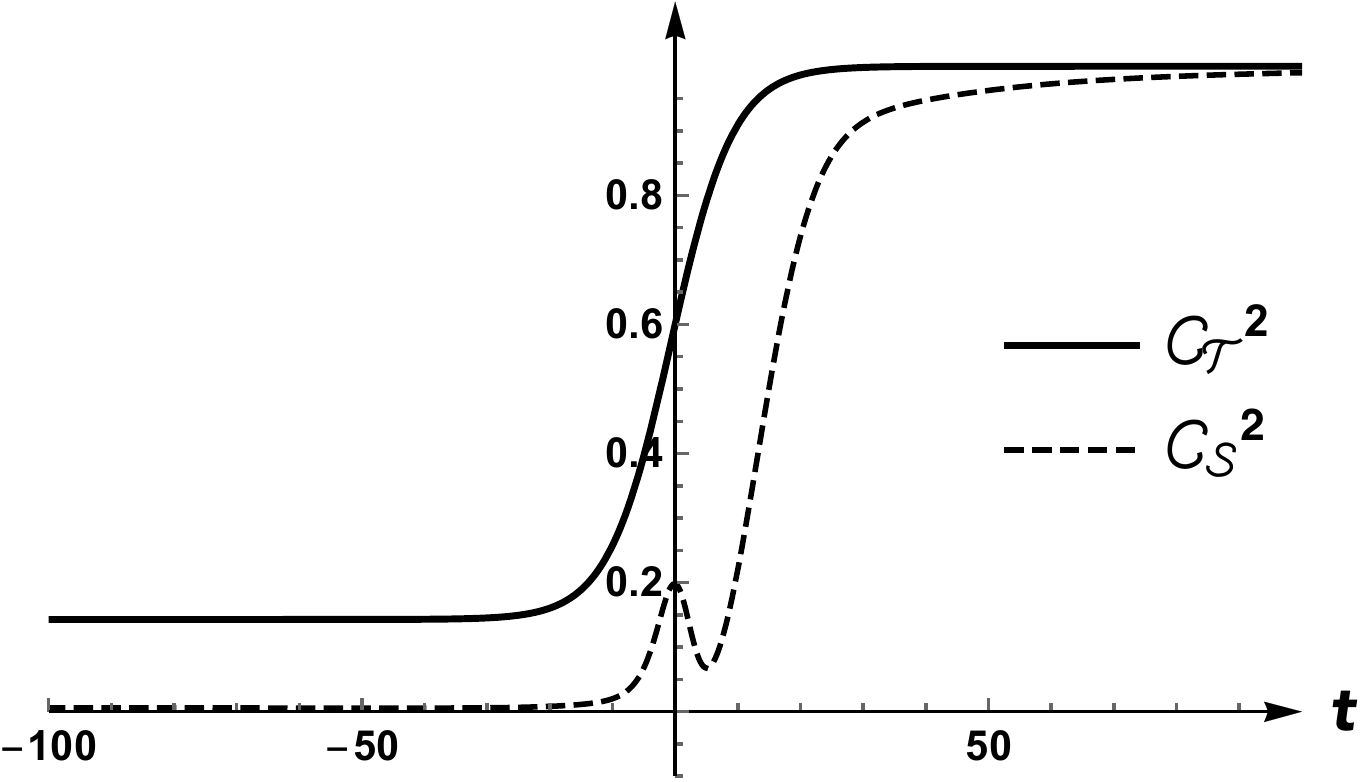}} (b) \\
\end{minipage}
\vfill\vspace{0.5cm}
\begin{minipage}[h!]{0.49\linewidth}
\center{\includegraphics[width=1\linewidth]{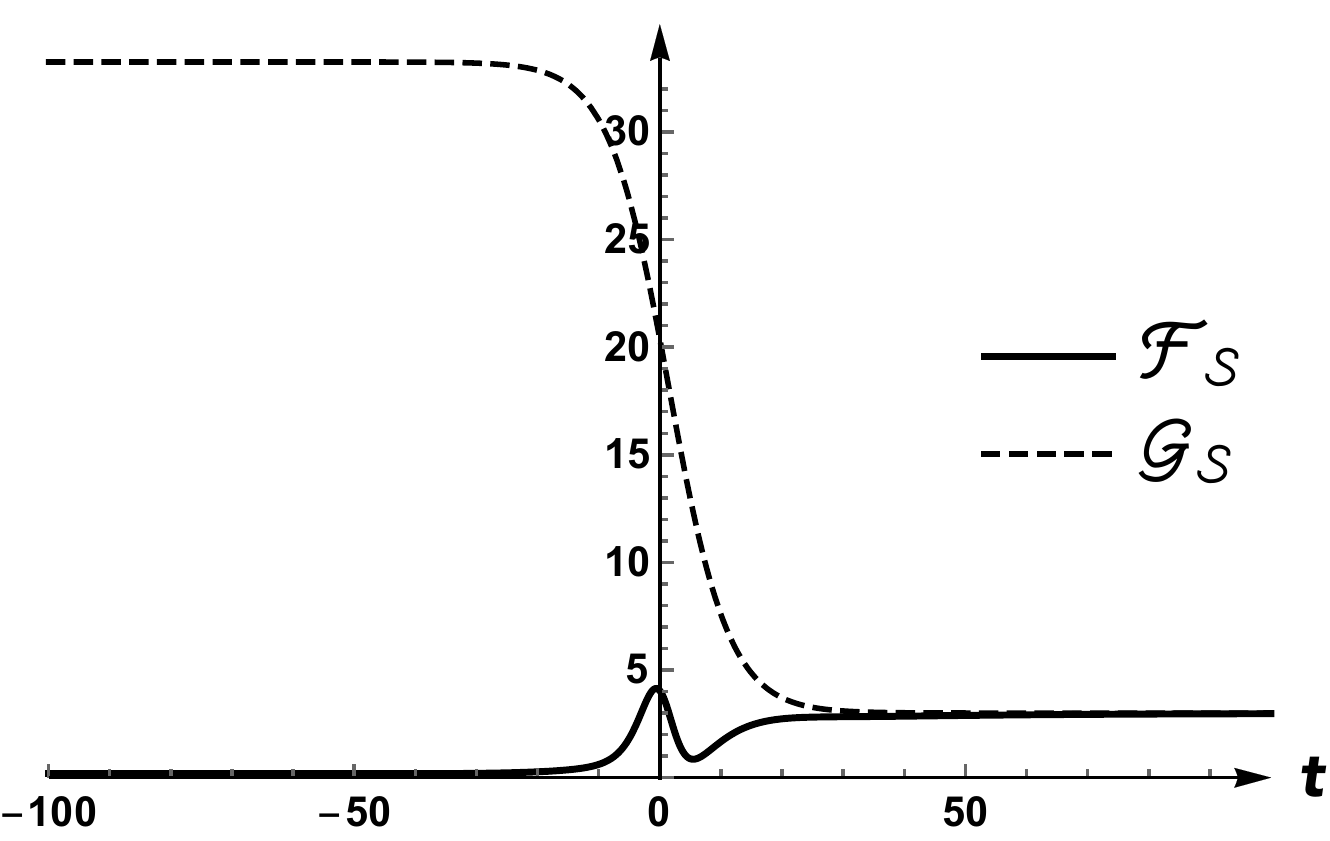}} (c) \\
\end{minipage}
\hfill
\begin{minipage}[h!]{0.49\linewidth}
\center{\includegraphics[width=1\linewidth]{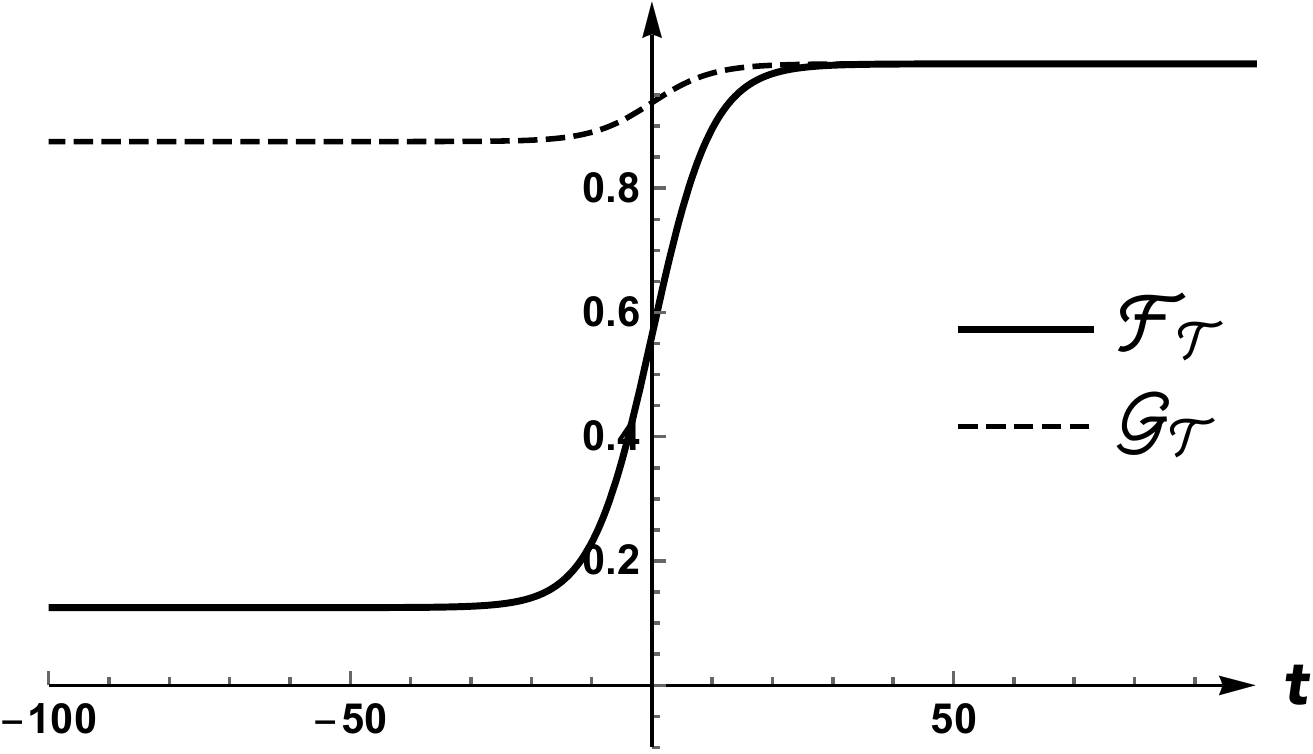}} (d) \\
\end{minipage}
\caption{ (a) The plots of $\xi$,
$(\mathcal{G_T}+\mathcal{D}\dot{\pi})$ and $\Theta$ for the model of 
Ref.~\cite{bouncegen}: $\xi$ crosses zero at $t \approx -1.039$
due to the behaviour of $(\mathcal{G_T}+\mathcal{D}\dot{\pi})$;
$\Theta$ is always positive, i.e., there is no $\gamma$-crossing.
(b) Sound speeds squared for the scalar and tensor modes:
$c_{\mathcal{S}}^2 \rightarrow 0.006$,
$c_{\mathcal{T}}^2 \to 0.18$
as $t\rightarrow -\infty$; $c_{\mathcal{S}}^2, c_{\mathcal{T}}^2 \to 1$
as $t\rightarrow +\infty$.
(c) The coefficients $\mathcal{G_S}$ and $\mathcal{F_S}$;
both are finite as $t\rightarrow -\infty$: $\mathcal{F_S}
\rightarrow 0.193$. 
(d) The coefficients $\mathcal{{G}_T}$ and $\mathcal{F_T}$.}
\label{pic:bouncegen_no-go}
\end{figure}
Since in this scenario $\Theta > 0$ at any time, the no-go theorem 
is circumvented by making a judicial choice for the function $F_4(\pi,X)$,
which determines the behaviour of $\mathcal{D}$ 
in eq.~\eqref{eq:xi_func_setup}: we have 
$(\mathcal{G_T} + \mathcal{D}\dot{\pi}) < 0$
as $t\to-\infty$ and $(\mathcal{G_T} + \mathcal{D}\dot{\pi}) > 0$
as $t\to+\infty$.
In Fig.~\ref{pic:bouncegen_no-go} (a) we plot $\xi$,
$(\mathcal{G_T} + \mathcal{D}\dot{\pi})$ and $\Theta$ in this scenario 
to illustrate that the key coefficient $\xi$ entering the would-be
 no-go theorem
is indeed a monotonously growing function and that it crosses zero together
with $(\mathcal{G_T} + \mathcal{D}\dot{\pi})$.
As shown in Fig.~\ref{pic:bouncegen_no-go}
(c) and (d), coefficients
$\mathcal{G_T}$, $\mathcal{F_T}$,
$\mathcal{G_S}$ and $\mathcal{F_S}$ in the quadratic 
action~\eqref{eq:unconstrained_action} are positive at all times,
hence there are neither ghosts nor gradient instabilities.

The sound speeds squared in the scalar and tensor sectors are shown
in Fig.~\ref{pic:bouncegen_no-go} (b). Both speeds are always positive 
and tend to the speed of light as $t\rightarrow +\infty$ in full accordance
with the required asymptotic behaviour described by GR with a massless scalar field.

Finally, let us give the asymptotic form of the Lagrangian as
$t \to\pm\infty$. As required, at late times,
 the Lagrangian has a simple form 
 of GR with
a massless scalar field:
\begin{equation}
\label{futureasymp_lagrangian}
\mathcal{L}|_{t=+\infty}=-\dfrac{1}{2}R + \dfrac{1}{3}\dfrac{(\partial\pi)^2}{\pi^2}=-\dfrac{1}{2}R + \dfrac{1}{2}(\partial\phi)^2 \; .
\end{equation}
At early times the
bouncing model with no $\gamma$-crossing is described by the Lagrangian
of beyond Horndeski type:
\begin{equation}
\label{pastasymp_lagrangian}
\begin{aligned}
\mathcal{L}|_{t=-\infty}
&=\mathcal{C}_0\cdot\dfrac{1}{\pi^2} + \left(\dfrac{1}{3} + \mathcal{C}_1\right)\dfrac{(\partial\pi)^2}{\pi^2}  + \mathcal{C}_2\dfrac{(\partial\pi)^4}{\pi^2}
+2\dfrac{(\partial\pi)^2}{\pi}\square\pi -  \frac{1}{16}(\partial\pi)^2 R\\
&+ \frac{1}{8}\left[(\square\pi)^2 - \nabla^{\mu\nu}\pi\nabla_{\mu\nu}\pi\right]
+ \frac12\epsilon^{\mu\nu\rho\sigma}{\epsilon^{\mu'\nu'\rho'}}_\sigma\nabla_\mu\pi\nabla_\mu'\pi\nabla_{\nu\nu'}\pi\nabla_{\rho\rho'}\pi ,
\end{aligned}
\end{equation}
where $\mathcal{C}_0 = 2.43$, $\mathcal{C}_1 = -5.53$ and
$\mathcal{C}_2 = 1.06$ are model dependent constants.
The theory~\eqref{pastasymp_lagrangian} does not reduce to GR,
in full accordance with the absence of
$\gamma$-crossing.


\subsubsection {Cosmological bounce with $\gamma$-crossing and
  two simple asymptotics}
 \label{subsec:two}

The bouncing model with $\gamma$-crossing~\cite{bounceI} is a modification
of the scenario discussed above. The main difference between the
two constructions is that in the present
case, there is $\gamma$-crossing happening at some moment of time.
This  allows to construct the solution with both
asymptotics $t \to + \infty$ and
$t \to - \infty$
described by the Lagrangian~\eqref{futureasymp_lagrangian}.

The Hubble parameter in this scenario coincides with that in the 
previous model, see eq.~\eqref{eq:hubble_bouncegen}. Since we aim to have
GR in both asymptotics  $t \to \pm \infty$, for the sake of simplicity
we choose $G_4(\pi,X)$ and $F_4(\pi,X)$ in the Lagrangian in such a way  that
$\mathcal{G_T}=\mathcal{F_T}=1$ during entire evolution.
Thus, there are no instabilities in the tensor sector, while
the gravitational waves always propagate at the speed of light,
$c_{\mathcal{T}}^2 = 1$.
\begin{figure}[h!]
\centering
\includegraphics[width=8.5cm,clip]{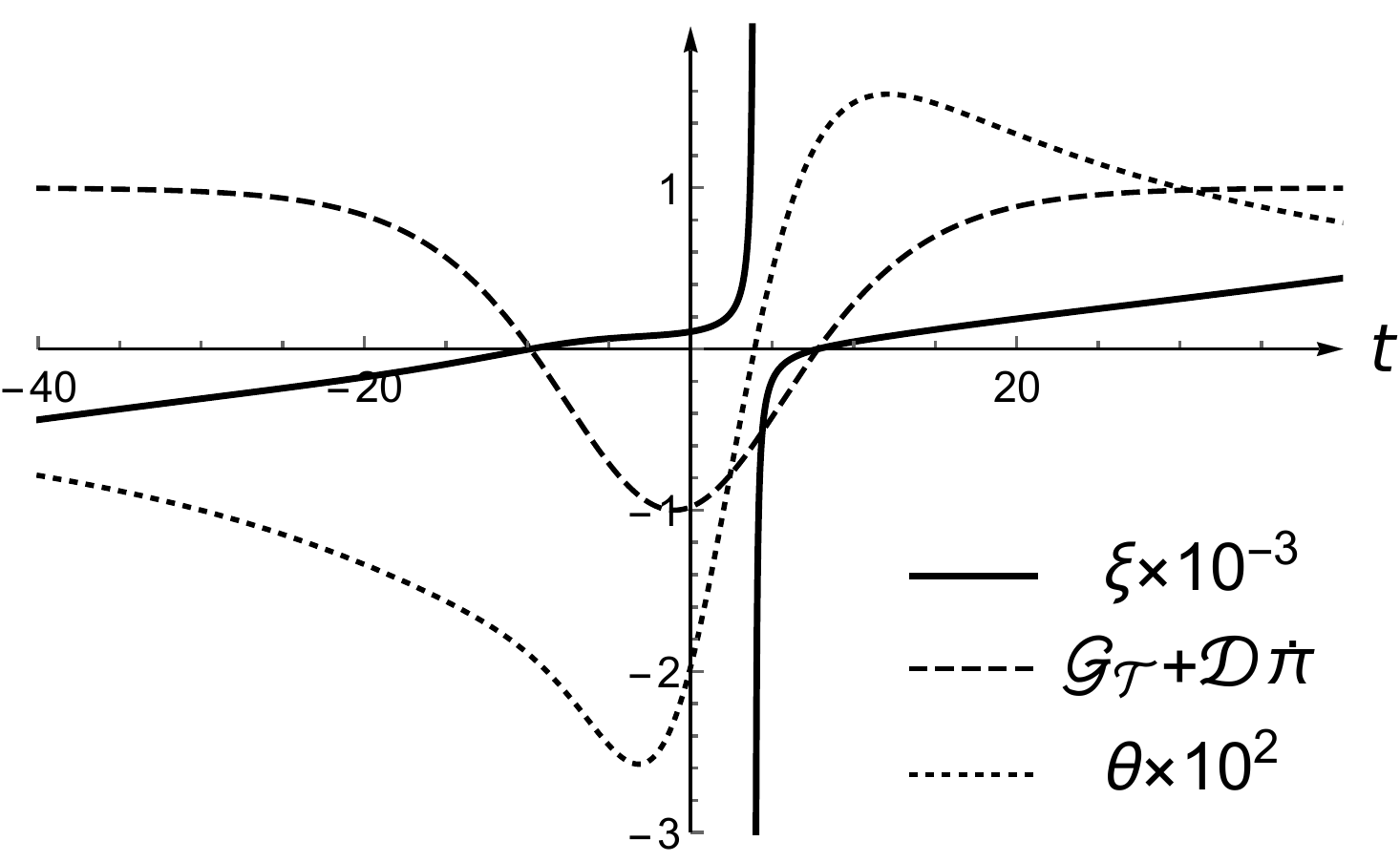}
\includegraphics[width=8.5cm,clip]{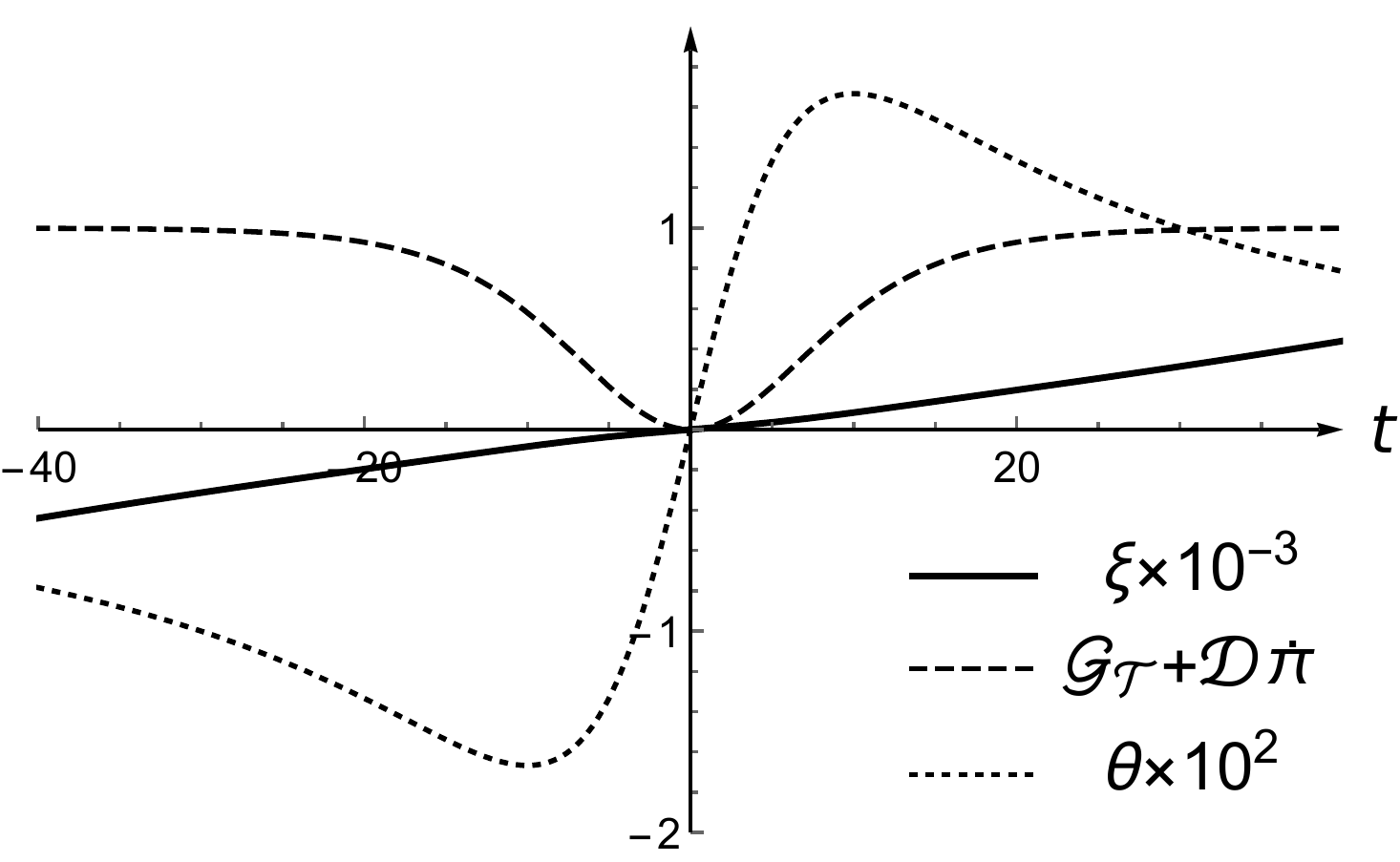}
\caption{The plots of $\xi$,
$(\mathcal{G_T}+\mathcal{D}\dot{\pi})$ and $\Theta$
for the model of Ref.~\cite{bounceI} (left panel):
$\xi$ crosses zero twice due to behaviour of
$(\mathcal{G_T}+\mathcal{D}\dot{\pi})$; $\Theta$ crosses zero and 
changes sign. The right panel illustrates the fine-tuned case:
$\xi$ remains finite at $\gamma$-crossing.}
\label{pic:bounceI_no-go}
\end{figure}

As discussed in Sec.~\ref{sec:gamma}, $\gamma$-crossing (sign change of 
$\Theta$) enables one to choose 
the Lagrangian function $F_4$ in such a way that 
$\mathcal{D}|_{t\to\pm\infty} \to 0$
(as before, $G_5=F_5=0$) and at the same time satisfy
the inequality $\dot{\xi} > \eps >0$ in eq.~\eqref{nogo} and
ensure that the scalar sector is free of gradient instabilities.
According to Fig.~\ref{pic:bounceI_no-go}, the function
$(\mathcal{G_T}+\mathcal{D}\dot{\pi})$ crosses zero twice,
while $(\mathcal{G_T}+\mathcal{D}\dot{\pi})|_{t\to\pm\infty} \to 1$, 
in full agreement with the choice
$\mathcal{G_T}=1$ at all times in this scenario.
As before, $\xi$ vanishes simultaneously with 
$(\mathcal{G_T}+\mathcal{D}\dot{\pi})$. Unlike in the
previous model, the reason for the negative sign of $\xi$ as
$t\to-\infty$ is negative $\Theta$: since the theory
reduces to GR, we have $\Theta \to H$ as $t \to \pm \infty$.
For completeness we illustrate the case of fine-tuned solution in 
Fig.~\ref{pic:bounceI_no-go} (right panel): despite
$\gamma$-crossing, $\xi$ is finite at all times
because
$(\mathcal{G_T}+\mathcal{D}\dot{\pi})$ touches zero
at the moment when $\Theta =0$.

\begin{figure}[h!]
\centering
\includegraphics[width=8.cm,clip]{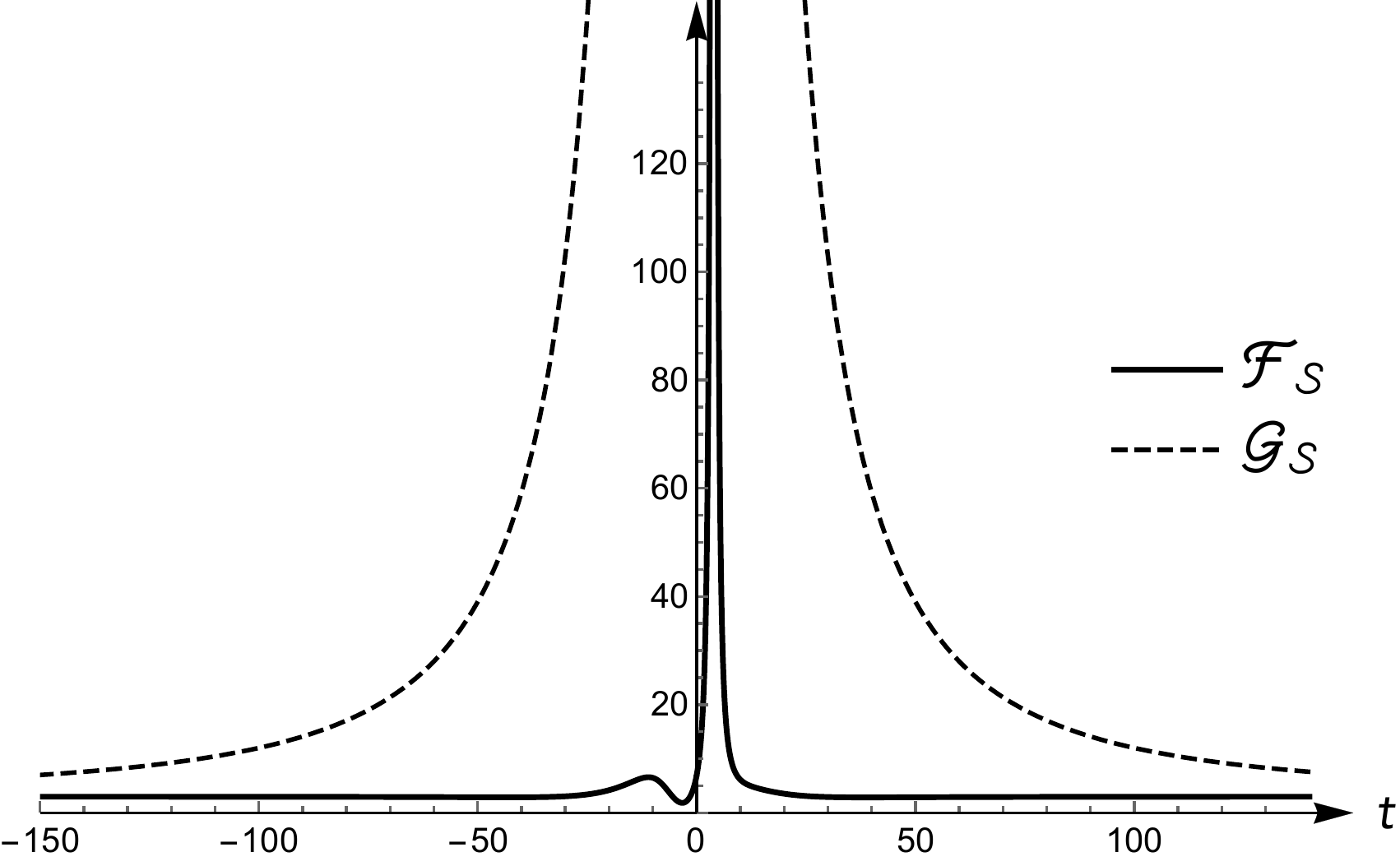}
\includegraphics[width=8.cm,clip]{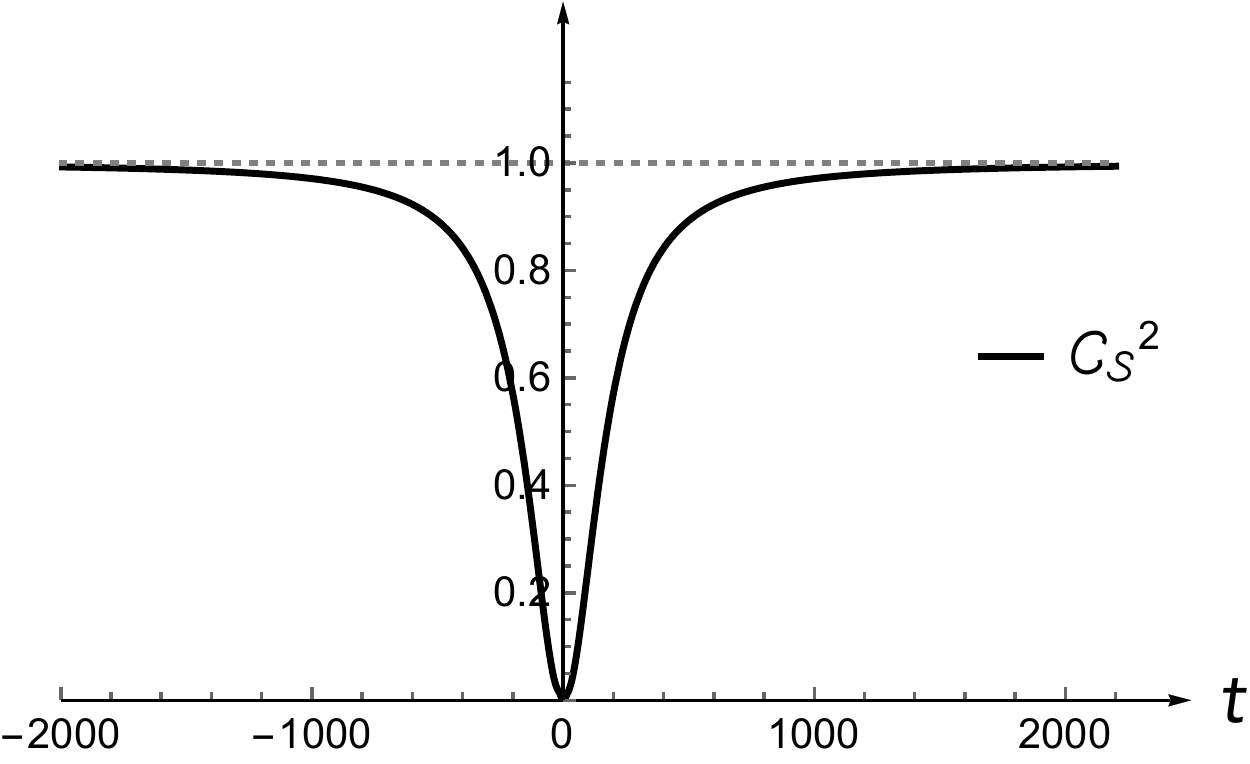}
\caption{The coefficients $\mathcal{G_S}$ and $\mathcal{F_S}$
(left panel):
both  are positive and diverge at the moment of 
$\gamma$-crossing, while their ratio is finite, strictly positive and
equals to the sound speed squared of the scalar mode (right panel);
$\min(c_{\mathcal{S}}^2) \simeq 0.001$.}
\label{pic:FsGs}
\end{figure}

To confirm the stability of the scalar sector, we show the 
functions $\mathcal{G_S}$ and $\mathcal{F_S}$ in Fig.~\ref{pic:FsGs}:
both coefficients are positive and diverge at the moment of 
$\gamma$-crossing, while their ratio is finite and strictly positive,
 in line with eqs.~\eqref{eq:GS_setup} 
and~\eqref{eq:FS_setup}.


\subsubsection{Genesis and its modifications in Horndeski theory and beyond}
 \label{subsec:three}

In the same paper~\cite{LMR}, where the no-go theorem was 
initially proven for the cubic Horndeski 
theory~\eqref{cubicH}, a modified version of the Genesis 
scenario evading the no-go theorem in the same cubic subclass was suggested:
the scale factor, instead of staying asymptotically constant,
tends to zero as $t\to -\infty$ (but in such a way that
the space-time curvature also vanishes). This enables one to avoid
gradient instabilities throughout
entire evolution. The price to pay for  complete 
stability, however, is geodesic incompleteness of the solution as $t\to-\infty$.
Unlike the original Genesis scenario of Ref.~\cite{gen1}
with $H \propto (-t)^{-3}$, in the modified version 
the Hubble parameter and scale factor during the Genesis-like stage have the
following time-dependence:
\be
\label{hubbleLMR}
H = -\frac{h}{t}, \qquad a(t)\propto\frac{1}{(-t)^h}, \qquad h=\const,
\qquad h>1 , \qquad t < 0,
\ee
while the energy density $\rho$ and effective pressure
$p$ behave as $t^{-2}$ as
$t\to-\infty$ (within the original scenario,
$\rho \propto t^{-6}$,
$p \propto t^{-4}$).

Later on, the analog of the modified Genesis was constructed 
in beyond Horndeski theory in Ref.~\cite{bouncegen}.
The Hubble parameter and the scale factor were chosen as
follows:
\be
\label{hubbleGenesis_like}
H(t) = \dfrac{1}{3\sqrt{\tau^2 + t^2}}, \qquad a(t) = \left[t +\sqrt{\tau^2 + t^2}\right]^\frac{1}{3},
\ee
where $\tau$ is a characteristic time scale.
The essentially new property here is sufficiently 
slow evolution of the scale factor,
$a(t) \propto |t|^{-1/3}$ as $t \to -\infty$,
and hence geodesic completeness.
The Genesis-like solution~\eqref{hubbleGenesis_like} evades
the no-go theorem in a similar way as  in the
above bouncing solution without $\gamma$-crossing. Here $\gamma$-crossing
is also forbidden, so the asymptotic behaviour of the theory
as $t\to-\infty$ corresponds to a substantially modified gravity of
beyond Horndeski type. Like in
the geodesically incomplete case~\eqref{hubbleLMR},
the theory transforms, as $t\to+\infty$, to GR +
massless scalar field~\eqref{futureasymp_lagrangian}.

Finally, in Ref.~\cite{genesisGR} a completely stable Genesis 
scenario with simple form of both asymptotics
was constructed within beyond Horndeski theory.
At early times the theory coincides with the original 
Genesis~\cite{gen2}:
\be
\label{hubble_genesisGR_past}
t \to -\infty: \quad
H = \dfrac{f^3}{4\Lambda^3} \dfrac{\left(1+\frac{\alpha}{3}\right)}{|t|^3} , \quad
a(t) = 1 + \dfrac{f^3}{8\Lambda^3} \dfrac{\left(1+\frac{\alpha}{3}\right)}{|t|^2},
\ee
where $\Lambda$, $f$ and $ \alpha$ are the same parameters as 
in Ref.~\cite{gen2}.
During the Genesis epoch the Lagrangian belongs to the cubic subclass
of Horndeski theories:
\be
\label{lagr_asymp_early}
\mathcal{L}_{t\to-\infty} = -\dfrac12 R -\dfrac{3 f^3}{4 \Lambda^3} \dfrac{(1+\alpha)}{\pi^4} \cdot X
+ \dfrac{3 f^3}{4 \Lambda^3} \dfrac{(1+\frac{\alpha}{3})}{\pi^4} \cdot X^2
- \dfrac{f^3}{2 \Lambda^3} \dfrac{X}{\pi^3} \cdot \Box\pi
,
\ee
and upon field redefinition
$\phi = f\cdot \log\left(-\sqrt{\frac{3f}{2\Lambda^3}}\frac{1}{\pi}\right)$,
its Lagrangian
coincides with that in Ref.~\cite{gen2}. 
Due to $\gamma$-crossing, the late time asymptotics of the theory
is GR: the Lagrangian 
transforms,  as $t\to+\infty$,
to the standard form~\eqref{futureasymp_lagrangian}, 
while $H = (3t)^{-1}$.

\begin{figure}[h!]
\centering
\includegraphics[width=8.cm,clip]{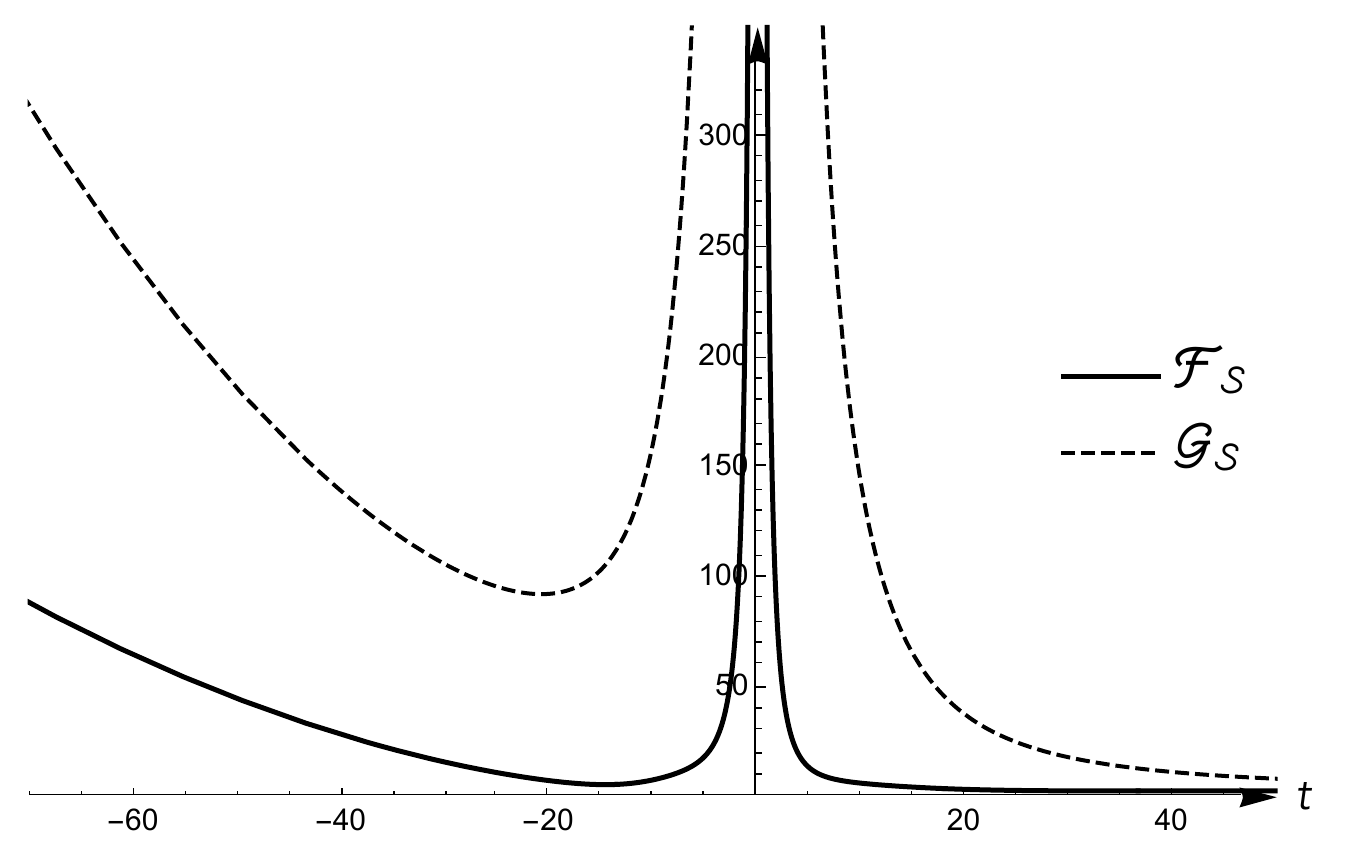}
\includegraphics[width=8.cm,clip]{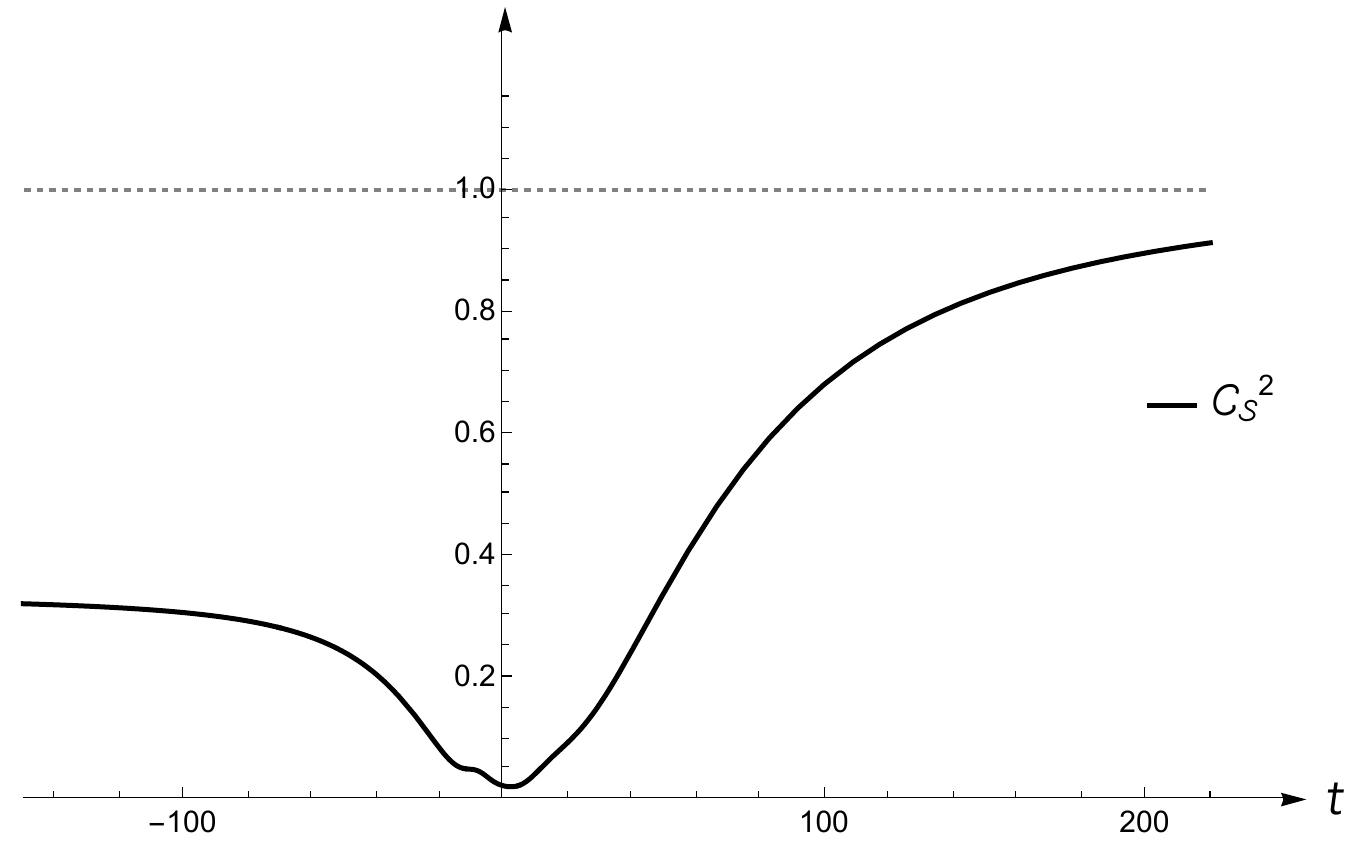}
\caption{The coefficients $\mathcal{G_S}$ and $\mathcal{F_S}$
  in the Genesis model \eqref{hubble_genesisGR}
(left panel):
both are positive and diverge at the moment of 
$\gamma$-crossing, while their ratio is finite, strictly positive and
equals to the sound speed squared of scalar mode (right panel);
$\min(c_{\mathcal{S}}^2) \simeq 0.02$.}
\label{pic:FsGs_genesisGR}
\end{figure}
The Hubble parameter in this model 
reads
\be
\label{hubble_genesisGR}
H(t) = \left[\left(
4\; \frac{\Lambda^3}{f^3} \cdot \frac{1- \mbox{tanh}(t/\tau)}{2\left(1+\alpha/3\right)} + 3 \cdot \frac{1 + \mbox{tanh}(t/\tau)}2
\right)\sqrt{2\tau^2 +t^2}\right]^{-1}.
\ee
The reconstruction procedure is analogous to that used
in the bouncing
case with $\gamma$-crossing:
$\mathcal{G_T}=\mathcal{F_T} =1$ at all times, so that 
tensor modes always propagate at the speed of light
($c_{\mathcal{T}}^2=1$); the behaviour of the key functions
entering the would-be no-go argument, namely, $\xi$,
$(\mathcal{G_T}+\mathcal{D}\dot{\pi})$ and $\Theta$, is similar
to that shown in
Fig.~\ref{pic:bounceI_no-go} 
(left panel).
Time-dependence of the coefficients $\mathcal{G_S}$ and $\mathcal{F_S}$
 responsible for stability of the scalar sector, is
 shown in Fig.~\ref{pic:FsGs_genesisGR}, where
$\mathcal{G_S}|_{t\to+\infty} \to 3$, 
$\mathcal{F_S}|_{t\to+\infty} \to 3$, and 
$c_{\mathcal{S}}^2|_{t\to+\infty} \to 1$.
At early times ($t\to -\infty$) we have
$\mathcal{G_S}|_{t\to-\infty}  \propto |t|^2$
and $\mathcal{F_S}|_{t\to-\infty}  \propto |t|^2$,
which is a distinguishing feature of the Genesis scenario,
while $c_{\mathcal{S}}^2|_{t\to -\infty} < 1$.


\section{Disformal transformation and no-go theorem}
\label{sec:disformal}
It was mentioned in Introduction that some of the subclasses of DHOST
and U-degenerate theories are
related to the Horndeski theories~\cite{DHOST6,DHOST7,disf1,disf2}
by the invertible disformal transformation~\eqref{intro:disformal}
of metric~\cite{disformal}, which is a generalization of 
the standard conformal transformation 
$\bar{g}_{\mu\nu} = \Omega^2(\pi) g_{\mu\nu}$.
Let us recall several features of disformal transformations.
Disformal transformations~\eqref{intro:disformal},
with $\Omega^2(\pi)$ and $\Gamma(\pi)$ being the functions of 
the scalar field $\pi$ only, do not extend the Horndeski subclass
$\mathcal{L}_2+\mathcal{L}_3+\mathcal{L}_4$ to a wider one.
The invertible transformations with 
$\Omega^2(\pi,X)=1$ and arbitrary $\Gamma(\pi,X)$
 extend the Horndeski theory 
$\mathcal{L}_2+\mathcal{L}_3+\mathcal{L}_4$
to its generalization with $F_4(\pi,X) \neq 0$~\cite{Zuma,disf0}.
Conversely,  the Lagrangian 
\begin{multline}
\label{lagrangian}
\mathcal{L}(F,K,G_4,F_4) = F(\pi,X) + K(\pi,X)\Box\pi - G_4(\pi,X)R \\
+ \left(2 G_{4X}(\pi,X) - F_4 (\pi,X) \; X\right)\left[\left(\Box\pi\right)^2-\pi_{;\mu\nu}\pi^{;\mu\nu}\right] \\
+ 2 F_4 (\pi,X) \left[\pi^{,\mu} \pi_{;\mu\nu} \pi^{,\nu}\Box\pi -  \pi^{,\mu} \pi_{;\mu\lambda} \pi^{;\nu\lambda}\pi_{,\nu} \right],
\end{multline}
which admits the stable solutions discussed in the previous section,
can be transformed by  the disformal transformation to the form  
$\mathcal{\bar{L}}=\mathcal{L}(\bar{F},\bar{K},\bar{G}_4)$, i.e., to the
''non-extended'' Horndeski theory (modulo a subtlety that we are about
to discuss).

One may wonder
whether there is a contradiction between the existence of 
completely stable solutions
in beyond Horndeski theories~\eqref{lagrangian} and the no-go theorem
valid in any Horndeski theory with $F_4(\pi,X) = 0$,
given  that these theories are apparently related by field redefinition.
The resolution of this ``paradox'' was given within the EFT approach in 
Ref.~\cite{Creminelli}: the pertinent disformal transformation 
is singular  right at the moment
when $\xi$  in eq.~\eqref{eq:xi_func_setup} crosses zero.
The latter fact was established in
Ref.~\cite{Creminelli} by analyzing the effective
action for perturbations written in  
the most general form. This section aims at obtaining
this result in the covariant framework. 

Let us consider the disformal transformation
\be
\label{disf}
\bar{g}_{\mu\nu} = g_{\mu\nu} + \Gamma_4(\pi,{X}) \partial_{\mu}\pi\partial_{\nu}\pi,
\ee
which converts the Lagrangian~\eqref{lagrangian} 
to the  Horndeski type:
\be
\label{BHtoH}
{\cal L}_4[{G}_4]  +   {\cal L}_4[{F}_4] =  \bar{\cal {L}}_4[\bar{G}_4].
\ee
The equation for the function $\Gamma_4(\pi,X)$, which implements the 
transformation~\eqref{BHtoH}, was found within the covariant
approach in Refs.~\cite{Gleyzes2,disf2}:
\be
\label{gammaX}
\Gamma_{4X} = \frac{F_4}{G_4 - 2 G_{4X}X - F_4 X^2} \,.
\ee
The relation between the new function $\bar{G}_4$ in Horndeski theory
${\cal \bar{L}}_4$ and the original $G_4$ in beyond Horndeski
theory ${\cal L}_4$ was established as well:
\be
\label{disf_G4}
\bar{G}_4 (\pi,\bar{X}) = \frac{G_4(\pi,X)}  {\sqrt{1 + X \Gamma_4}}\,,
\qquad \bar{X} = \frac{X}{1 + X \Gamma_4} \,.
\ee
Let us demonstrate that the transformed function $\bar{G}_{4\bar{X}}$
which enters the Lagrangian ${\cal \bar{L}}_4$
becomes singular when $\xi$ crosses zero together with 
$(\mathcal{G_T}+\mathcal{D}\dot{\pi})$ in eq.~\eqref{eq:xi_func_setup}.

The relation between the functions $\bar{G}_{4\bar{X}}$ and $G_{4X}$,
follows from the transformation 
rules~\eqref{disf_G4} and reads:
\be
\label{disf_G4X}
\bar{G}_{4\bar{X}} =
\frac{\partial \bar{G}_4}{\partial \bar{X}} =
\frac{\sqrt{1+X\Gamma_4}}{1-X^2\Gamma_{4X}}
\left( G_4 (1+X\Gamma_4)- \frac{1}{2} G_{4}(\Gamma_4 + X \Gamma_{4X}) \right).
\ee
Let us recast $\Gamma_{4X}$ in eq.~\eqref{gammaX} in terms of
notations in eqs.~\eqref{eq:GT_coeff_setup} and~\eqref{eq:D_coeff_setup}:
\be
\label{gammaX_transf}
\Gamma_{4X} = \frac{\mathcal{D}\dot{\pi}}{X^2(\mathcal{G_T}+2\mathcal{D}\dot{\pi})} \,.
\ee
We now substitute the expression for $\Gamma_{4X}$
in eq.~\eqref{gammaX_transf}
into the denominator factor in eq.~\eqref{disf_G4X}:
\be
\label{transf_flaw}
\frac{1}{1-X^2 \Gamma_{4X}} =
\frac{(\mathcal{G_T}+2\mathcal{D}\dot{\pi})}{\mathcal{G_T}+\mathcal{D}\dot{\pi}}.
\ee
It follows from eq.~\eqref{transf_flaw} that the denominator of 
the transformation~\eqref{disf_G4X} crosses zero right
at the moment when $\xi=0$, see eq.~\eqref{eq:xi_func_setup},
so that $\bar{G}_{4\bar{X}}$ is divergent at that moment of time.
According to the discussion in Sec.~\ref{sec:no-go}, one
evades the no-go theorem by going beyond Horndeski and having
$\mathcal{D} \neq 0$.
This enables one to satisfy the requirements for $\xi$ 
given by eq.~\eqref{nogo}. The latter imply that
$\mathcal{G_T}+\mathcal{D}\dot{\pi}$ vanishes at some moment(s) of time.
Therefore, the beyond Horndeski theories which admit 
completely stable non-singular solutions, are disformally 
``related'' to the Horndeski theories by singular transformations.
So there is, in fact, no contradiction between the existence of completely
stable solutions and the no-go theorem in apparently disformally
related theories.


\section{Conclusion.}
\label{sec:conclusion}

In this mini-review we have briefly discussed the recent studies
of non-singular cosmological scenarios and their stability 
in beyond Horndeski theory.

We have described specific examples of bouncing Universe and Genesis
models in beyond Horndeski theory, which are free of ghost and 
gradient instabilities during entire evolution 
from $t\to-\infty$ to $t\to+\infty$. A nice feature 
of some of these models is the simple form of the theory in the 
asymptotics $t\to\pm\infty$, which is GR with a 
conventional massless scalar field. The advantage of having
the asymptotic behaviour described by GR becomes clear when
attempting to construct reasonably realistic bouncing and Genesis scenarios. 
In particular,
the models make use of the Galileon property
of safe NEC/NCC violation, crucial for
the bounce and Genesis, while they avoid dealing with exotic matter
away from the NEC/NCC-violating regime.  
Although the analysis of
phenomenological prospects of these cosmological scenarios
is  beyond the scope of this review, we think this is a promising
field of research.

A particular topic we have addressed is
the disformal relation between
beyond Horndeski and  Horndeski theories.
Our motivation here was to
collect
results obtained in the covariant formalism so far. 
The question about the consistency 
of the no-go theorem in Horndeski theory, on the one hand,
and the existence of completely stable solutions in
beyond Horndeski theory, on the other, in view of their
relation by field redefinition, has been often raised even
after the appearance
of Ref.~\cite{Creminelli}.
We consider
our confirmation, in the covariant formalism,
of the results of Ref.~\cite{Creminelli} 
a useful addition
which fully resolves the issue.


\section{Acknowledgements}
The authors are grateful to R. Kolevatov and N. Sukhov for 
productive collaboration and to E.~Babichev and A.~Vikman 
for helpful discussions. This work has been supported by the
Russian Science Foundation grant 19-12-00393.

\end{document}